\documentclass[prd,showpacs,floatfix,amsmath,amssymb]{revtex4}
\usepackage{bm}
\usepackage[dvips,bookmarks]{hyperref}
\usepackage{graphicx}
\usepackage{dcolumn}
\newcommand{\ud}{\text{d}} 
\begin{document}

\title{Rates of Neutrino Absorption on Nucleons and 
the Reverse Processes in Strong Magnetic Fields}

\author{Huaiyu Duan}

\thanks{Present address: Center for Astrophysics \& Space Sciences, 
University of California, San Diego, La Jolla, CA 92093-0424}

\email{hduan@ucsd.edu}

\author{Yong-Zhong Qian}

\email{qian@physics.umn.edu}

\affiliation{School of Physics and Astronomy, 
University of Minnesota, Minneapolis,
MN 55455}

\date{\today}

\begin{abstract}

The rates of $\nu_e + n \rightleftharpoons e^- + p$ and
$\bar\nu_e + p \rightleftharpoons e^+ + n$ are important
for understanding the dynamics of supernova explosion and
the production of heavy elements in the supernova environment 
above the protoneutron star. Observations and theoretical 
considerations suggest that some protoneutron stars may be 
born with strong magnetic fields. In a previous paper we
calculated the above rates in supernova environments 
with magnetic fields up to $\sim 10^{16}$~G assuming that
the nucleon mass $m_N$ is infinite. We also applied these rates 
to discuss the implications of such strong fields for supernova
dynamics. In the present paper we take into account the effects
of a finite $m_N$ and develop a numerical method to recalculate 
the above rates in similar environments. This method 
is accurate to $\mathcal{O}(1/m_N)$ and such an accuracy is
required for application to supernova nucleosynthesis. We
show that our results have the correct behavior in the
limit of high neutrino energy or small magnetic field.
Based on comparison of our results with various 
approximations, we recommend efficient estimates of the 
above rates for use in models of supernova nucleosynthesis 
in the presence of strong magnetic fields.

\end{abstract}
\maketitle

\section{Introduction}

The processes
\begin{subequations}
\label{eq:processes}
\begin{eqnarray}
\nu_{e}+n & \rightleftharpoons & e^{-}+p,\label{eq:nun}\\
\bar{\nu}_{e}+p & \rightleftharpoons & e^{+}+n,\label{eq:nup}
\end{eqnarray}
\end{subequations}
play important roles in supernovae. A supernova is initiated by
the collapse of a stellar core, which leads to the formation of a
protoneutron star. Nearly all the gravitational binding energy 
of the protoneutron star is emitted in $\nu_{e}$, 
$\bar{\nu}_{e}$, $\nu_{\mu}$, $\bar{\nu}_{\mu}$, $\nu_{\tau}$, 
and $\bar{\nu}_{\tau}$, some of which would interact to heat the 
material above the protoneutron star. The forward neutrino 
absorption processes in Eq.~\eqref{eq:processes} provide the
dominant heating mechanism, which is counteracted by cooling of
the material through the reverse neutrino emission processes. 
In a prevalent paradigm 
\cite{Bethe:1985}, supernova explosion is determined by the 
competition between these heating and cooling processes. 
These processes also interconvert neutrons
and protons, thereby setting the neutron-to-proton ratio of the 
material above the protoneutron star \cite{Qian:1996}. This ratio
is a key parameter that governs the production of heavy elements 
during the ejection of this material \cite{Woosley:1992,Hoffman:1996}.
Thus, accurate rates of the processes in Eq.~\eqref{eq:processes} are
important for understanding supernova dynamics and nucleosynthesis.

Observations and theoretical considerations indicate that
protoneutron stars with magnetic fields of $\sim 10^{16}$ G
may be formed. The rates of the above processes in such
strong fields have been studied in the literature with various
approximations 
\cite{Leinson:1998,Roulet:1998sw,%
Lai:1998sz,Arras:1998mv,Gvozdev:1999,Chandra:2002,%
Bhattacharya:2002qf,Shinkevich:2004ja}.
In our previous work \cite{Duan:2004nc}, we used the Landau wave
functions of $e^\pm$ and derived a set of
simple and consistent formulas to calculate the rates of
the processes in Eq.~\eqref{eq:processes} 
in the presence of strong magnetic fields. We also applied these
rates to discuss the implications of such fields for supernova
dynamics. However, all the calculations in the literature, including 
our previous work, were mostly carried out to $\mathcal{O}(1)$, the 
zeroth order in $1/m_N$ with $m_N$ being the nucleon mass. None of
them included both the effects of nucleon recoil and weak magnetism, 
which are of $\mathcal{O}(1/m_{N})$ and known to be important
for the conditions in supernovae \cite{Horowitz:2001xf}. For 
modeling the production of heavy elements during the ejection of 
the material from the protoneutron star, an accuracy of $\sim 1$\%
for the rates of the processes in Eq.~\eqref{eq:processes} is
required to determine precisely the neutron-to-proton ratio in the 
material \cite{Hoffman:1996}. To achieve such an accuracy, the
$\mathcal{O}(1/m_{N})$ effects on these rates must be taken into
account. In this paper we recalculate these rates using the 
respective Landau wave functions of $e^\pm$ and protons and including 
the $\mathcal{O}(1/m_{N})$ corrections from both nucleon recoil and 
weak magnetism. Our goal is to identify the important factors in
computing accurate rates of the processes in Eq.~\eqref{eq:processes}
for application to supernova nucleosynthesis in the presence of 
strong magnetic fields.

This paper is organized as follows. The energies and 
the wave functions of the relevant particles in magnetic fields
are discussed in Sec.~\ref{sec:energies}. The cross 
sections of the neutrino absorption processes in 
Eq.~\eqref{eq:processes} and the differential 
reaction rates of the reverse neutrino emission processes 
are derived to $\mathcal{O}(1/m_{N})$ in Sec.~\ref{sec:rates}. 
The rates of these processes in supernova environments with 
magnetic fields of $\sim 10^{16}$ G are calculated and discussed
in Sec.~\ref{sec:results}. Conclusions are given in 
Sec.~\ref{sec:conclusions}.

\section{Particle energies and wave functions in magnetic fields
\label{sec:energies}}

The importance of magnetic field effects can be gauged from the energy
scale
\begin{equation}
\sqrt{eB}=7.69 \left(\frac{B}{10^{16}\,\text{G}}\right)^{1/2}\,\text{MeV},
\label{eq:scale}
\end{equation}
where $e$ is the charge of $e^+$ and $B$ is the field strength.
However, there is no detailed knowledge of magnetic fields in supernovae.
Observations indicate that neutron stars may have  
$B\sim 10^{15}\text{ G}$  long after their birth in supernovae
\cite{Kouveliotou:1999,Gotthelf:1999,Ibrahim:2003}. This suggests that 
at least $B\sim 10^{15}\text{ G}$ can be generated
during the formation of some protoneutron stars. A recent theoretical
model suggests that $B\sim 10^{16}\text{ G}$ may be produced near the
surface of a protoneutron star \cite{Akiyama:2003}. An upper 
limit of $B\sim 10^{18}\text{ G}$ can be estimated for such a star by 
equating the magnetic energy to its gravitational binding energy 
\cite{Lai:2000at}. To explore the effects of strong magnetic fields on
the rates of the processes in Eq.~\eqref{eq:processes}, we consider 
that $B\sim 10^{16}\text{ G}$ may exist in the region of interest to
supernova nucleosynthesis, which lies well below $10^7$~cm from the 
protoneutron star \cite{Qian:1996}. 
For such fields, the associated energy scale
is much smaller than the mass of the $W$ boson $M_W=80$ GeV. So
there will be no change in the description of the weak interaction that 
is involved in the processes in Eq.~\eqref{eq:processes}. On the other
hand, the energy scale for $B\sim 10^{16}\text{ G}$ is larger than the
temperature ($T\sim 1$~MeV) of the material above the protoneutron star
and comparable to the typical neutrino energy ($E_\nu\sim 10$~MeV). Thus,
magnetic field effects on energy levels of charged particles ($e^\pm$ 
and $p$) will be important. Furthermore, $B\sim 10^{16}\text{ G}$ 
will induce polarization of nucleon spin at the level of 
$eB/m_NT\sim 10^{-2}$. This is significant due to parity violation 
of weak interaction and should also be taken into account.

We discuss the energy levels and the corresponding wave functions
of all the relevant particles in this section. We assume a uniform 
magnetic field $\mathbf{B}$ in the positive \textit{z}-direction, 
for which the vector potential is
\begin{equation}
\mathbf{A} = (-\frac{1}{2}By, \frac{1}{2}Bx, 0).
\end{equation}
All the wave functions will be given in Dirac-Pauli representation.

\subsection{Electron and positron
\label{sec:ewave}}

The motion of $e^\pm$ along
the $z$-axis is not affected by the magnetic field, but the motion 
in the $xy$-plane is quantized into Landau levels with energies
(see, e.g., Ref.~\cite{Landau:1977})
\begin{equation}
E_e = \sqrt{m_e^2 + k_{ez}^2 + 2 n_e e B},
\label{eq:energy-e}
\end{equation}
where $m_e$ is the rest mass of $e^\pm$, $k_{ez}$ is the 
$z$-component of the 
momentum, and $n_e$ is an integer quantum number (i.e.,
$n_e=0,\,1,\,2,\,\dots$). For the $e^\pm$ in the initial states of
the neutrino emission processes in Eq.~\eqref{eq:processes}, the
relevant $E_e$ is of the order of the temperature $T\sim 1$~MeV for
the material above the protoneutron star. It can be seen from
Eqs.~\eqref{eq:scale} and \eqref{eq:energy-e} that these $e^\pm$
predominantly occupy the ground Landau level ($n_e=0$) for
$B\sim 10^{16}\text{ G}$. In comparison, the $e^\pm$ in the 
final states of the neutrino absorption processes typically
have $E_e$ of the order of the neutrino energy $E_\nu\sim 10$ MeV.
These $e^\pm$ can occupy excited Landau levels ($n_e\geq 1$).

The wave function of $e^{-}$ in cylindrical coordinates 
$(\xi, \phi, z)$ is
\begin{equation}
(\psi_{e^-})_{s_e} = 
\frac{e^{i (k_{ez} z - E_{e} t)} e^{i (n_e - r_e) \phi}}%
{\sqrt{2 \pi L /eB}}
(U_{e^-})_{s_e},
\label{eq:electron-wave}
\end{equation}
where $s_e=1$ and $-1$ for spin up and down, respectively,
$r_e$ is the quantum number labeling the center of gyromotion 
in the $xy$-plane,
and $L$ is the linear size of the normalization volume.
In Eq.~\eqref{eq:electron-wave}, the spinor $(U_{e^-})_{s_e}$ is
\begin{align}
(U_{e^-})_{s_e=1} &=
\frac{1}{\sqrt{2 E_e (E_e + m_e)}}
\left(	\begin{array}{c}
 	(m_{e} + E_{e}) e^{-i \phi} I_{n_e-1,r_e}(eB\xi^2/2)\\
 	0\\
 	k_{ez} e^{-i \phi} I_{n_e-1,r_e}(eB\xi^2/2)\\
 	i \sqrt{2n_e eB} I_{n_e,r_e}(eB\xi^2/2)
	\end{array}
\right)
\\
\intertext{and}
(U_{e^-})_{s_e=-1} &=
\frac{1}{\sqrt{2 E_e (E_e + m_e)}}
\left(	\begin{array}{c}
 	0\\
 	(m_{e} + E_{e}) I_{n_e, r_e}(eB\xi^2/2)\\
 	-i \sqrt{2n_e eB} e^{-i \phi} 
	I_{n_e-1, r_e}(eB\xi^2/2)\\
 	-k_{ez} I_{n_e, r_e}(eB\xi^2/2)
	\end{array}
\right).
\end{align}
The special function $I_{n,r}(\zeta)$ in the above equations 
is defined in Ref.~\cite{Sokolov:1968} and can be calculated using
the method given in Appendix~\ref{sec:algorithm-of-I}.

The wave function of $e^{+}$ is 
\begin{equation}
(\psi_{e^+})_{s_e} = 
\frac{e^{-i(k_{ez} z - E_{e} t)} e^{i (n_e - r_e)\phi}}%
{\sqrt{2 \pi L /eB}}
(U_{e^+})_{s_e},
\end{equation}
where
\begin{align}
(U_{e^+})_{s_e=1} &=
\frac{1}{\sqrt{2 E_e (E_e + m_e)}}
\left(	\begin{array}{c}
 	i \sqrt{2n_e eB} e^{-i \phi}
	I_{n_e-1, r_e}(eB\xi^2/2)\\
 	-k_{ez} I_{n_e, r_e}(eB\xi^2/2)\\
 	0\\
 	(m_{e} + E_{e}) I_{n_e ,r_e}(eB\xi^2/2)
	\end{array}
\right)
\\
\intertext{and}
(U_{e^+})_{s_e=-1} &=
\frac{1}{\sqrt{2 E_e (E_e + m_e)}}
\left(	\begin{array}{c}
 	-k_{ez} e^{-i \phi} I_{n_e-1, r_e}(eB\xi^2/2)\\
 	i \sqrt{2n_e eB} I_{n_e, r_e}(eB\xi^2/2)\\
 	-(m_{e} + E_{e}) e^{-i \phi} I_{n_e-1, r_e}(eB\xi^2/2)\\
 	0
	\end{array}
\right).
\end{align}

Clearly, $E_e$ does not depend on the quantum number $r_e$ in the
wave functions. This leads to a degeneracy factor
\begin{equation}
\sum_{r_e} 1 = \frac{eB L^2}{2\pi}
\label{eq:degeneracy-r}
\end{equation} 
for each Landau level of $e^\pm$ (see, e.g., Ref.~\cite{Lai:2000at}).
Each level is further degenerate with respect to spin
except for the ground level 
[$(\psi_{e^-})_{s_e=1}=(\psi_{e^+})_{s_e=-1}=0$ for $n_e=0$]. 
This introduces an additional spin 
degeneracy factor $g_{n_e}$, which is 1 for $n_e=0$ and 2 for 
$n_e\geq 1$.

\subsection{Proton
\label{sec:pwave}}

Protons are nonrelativistic in the supernova environment of interest.
For nonrelativistic $e^+$ with the same charge and spin as protons,
expansion of Eq.~\eqref{eq:energy-e} to $\mathcal{O}(1/m_{e})$ gives
\begin{equation}
E_{e,\text{NR}} = m_e+\frac{k_{ez}^2}{2m_e}+\frac{n_e e B}{m_e}.
\end{equation}
The above equation already accounts for the contribution
from the $e^+$ magnetic moment of $e/2m_e$. Unlike $e^+$, protons
have an anomalous magnetic moment of
$\tilde{\mu}_{p}=1.79\,\mu_N$ in addition to the value
$\mu_N = e / 2m_p$ expected for a spin-1/2 point particle of
charge $e$ and mass $m_p$. Taking this into consideration, we
obtain the energies of the proton Landau levels as
\begin{equation}
E_{p} = m_{p}  + \frac{k_{pz}^{2}}{2m_{p}}
 + \frac{n_{p}eB}{m_{p}}- s_{p} \tilde{\mu}_{p} B,
\label{eq:energy-p}
\end{equation}
where symbols have similar meanings to those for $e^\pm$. For
$B\sim 10^{16}\text{ G}$, $eB/m_p\sim \tilde{\mu}_{p}B\sim 60$ keV.
The protons in the initial states of the $\bar\nu_e$ absorption and 
$\nu_e$ emission processes in Eq.~\eqref{eq:processes} have 
$E_{p}-m_{p}\sim T\sim 1$~MeV, and therefore, can occupy many Landau 
levels. By the correspondence principle, the quantum effects of
the magnetic field on these protons are insignificant.
However, the protons in the final states of
the $\nu_e$ absorption and $\bar\nu_e$ emission processes are less
energetic, with typical recoil energies of 
$E_{p}-m_{p}\sim E_\nu^2/m_p\sim 100$~keV and $\sim T^2/m_p\sim 1$~keV,
respectively. Thus, proper treatment of proton Landau levels is
especially important for these processes.

The proton wave function can be written as
\begin{equation}
(\psi_{p})_{s_{p}} = \frac{e^{i(k_{pz}z-E_{p}t)} e^{i(r_{p}-n_{p})\phi}}{%
\sqrt{2\pi L /eB}} (U_{p})_{s_{p}},
\end{equation}
where
\begin{align}
(U_{p})_{s_{p}=1} &= \left(
\begin{array}{c}
I_{n_{p},r_{p}}(eB \xi^{2}/2)\\
0\\
(k_{pz}/2m_{p}) I_{n_{p},r_{p}}(eB \xi^{2}/2)\\
-i e^{i\phi} (\sqrt{2n_{p}eB}/2m_{p})
I_{n_{p}-1,r_{p}}(eB \xi^{2}/2)
\end{array}
\right)
\label{eq:proton-spinor-up}
\\
\intertext{and} 
(U_{p})_{s_{p}=-1} &= \left(
\begin{array}{c}
0\\
e^{i\phi} I_{n_{p}-1,r_{p}}(eB \xi^{2}/2)\\
i(\sqrt{2n_{p}eB}/2m_{p}) I_{n_{p},r_{p}}(eB \xi^{2}/2)\\
-e^{i\phi} (k_{pz}/2m_{p}) I_{n_{p}-1,r_{p}}(eB \xi^{2}/2)
\end{array}\right).
\label{eq:proton-spinor-down}
\end{align}
Note that each proton Landau level is also degenerate with respect to
the quantum number $r_p$, but the spin degeneracy of the excited levels 
is lifted due to the contribution from the anomalous magnetic moment.

The proton wave function contains terms of the form
\begin{equation}
\Psi_{n,r} \equiv \frac{e^{ik_{z}z} e^{i(r-n)\phi}}{%
\sqrt{2\pi L/eB}} I_{n,r}(eB \xi^{2}/2),
\end{equation}
which has the following properties:
\begin{subequations}
\label{eq:raising-lowering}
\begin{align}
\pi_{+}\Psi_{n-1,r} 
& \equiv (\pi_x - i \pi_y) \Psi_{n-1,r}
= i\sqrt{2neB} \Psi_{n,r},
\label{eq:raising-rel}
\intertext{and}
\pi_{-}\Psi_{n,r} 
& \equiv (\pi_x + i \pi_y) \Psi_{n,r}
= -i\sqrt{2neB} \Psi_{n-1,r}.
\label{eq:lowering-rel}
\end{align}
\end{subequations}
The operator $\bm{\pi}$ in the above equations is defined as
\begin{equation}
\bm{\pi} \equiv -i \bm{\nabla} - e \mathbf{A}.
\end{equation}
Equations~\eqref{eq:raising-rel} and \eqref{eq:lowering-rel}
can be used to simplify the evaluation of the transition amplitudes
for the processes in Eq.~\eqref{eq:processes} \cite{Duan:2004th}.

\subsection{Neutron}

Neutrons are also nonrelativistic in the supernova environment of 
interest, and their energy is
\begin{align}
E_{n} &= m_{n} + \frac{k_{n}^{2}}{2 m_{n}} - s_{n}\mu_{n} B,
\end{align}
where $\mu_{n}=-1.91\,\mu_N$ is the neutron
magnetic moment. The corresponding wave function to 
$\mathcal{O}(1/m_N)$ is
\begin{equation}
(\psi_{n})_{s_{n}} = 
\frac{e^{i(\mathbf{k}_{n}\cdot\mathbf{x}-E_{n}t)}}{L^{3/2}}
(U_{n})_{s_{n}},
\label{eq:neutron-wave}
\end{equation}
where
\begin{align}
(U_{n})_{s_{n}=1} &= \left(
\begin{array}{c}
1\\
0\\
(k_{n}/2m_{n}) \cos\Theta_{n}\\
(k_{n}/2m_{n}) \sin\Theta_{n} e^{i\Phi_{n}}
\end{array}
\right)
\\
\intertext{and}
(U_{n})_{s_{n}=-1} &= \left(
\begin{array}{c}
0\\
1\\
(k_{n}/2m_{n}) \sin \Theta_{n}e^{-i\Phi_{n}}\\
-(k_{n}/2m_{n}) \cos \Theta_{n}
\end{array} \right).
\end{align}
In the above equations, $\Theta_n$ and $\Phi_n$ are the polar and 
azimuthal angles of the neutron momentum $\mathbf{k}_{n}$ in 
spherical coordinates. 

\subsection{Neutrinos}

The neutrino energy is not affected by the magnetic field.
For left-handed $\nu_{e}$ with momentum $\mathbf{k}_{\nu}$, 
the wave function is
\begin{equation}
\psi_{\nu_e} = 
\frac{e^{i (\mathbf{k}_{\nu} \cdot \mathbf{x}-E_{\nu } t )}}{L^{3/2}}
U_{\nu},
\end{equation}
where
\begin{equation}
U_{\nu}=\left(\begin{array}{c}
 	\sin (\Theta_{\nu}/2)\\
 	-\cos (\Theta_{\nu}/2)\\
	 -\sin (\Theta_{\nu}/2)\\
 	\cos (\Theta_{\nu}/2)
	\end{array}
\right).\label{eq:U-nu}
\end{equation} 
The azimuthal angle of $\mathbf{k}_{\nu}$ is taken to be $\Phi_\nu=0$ 
in the above equation. The wave function of right-handed $\bar\nu_e$ 
with the same momentum $\mathbf{k}_{\nu}$ is
\begin{equation}
\psi_{\bar\nu_e} = 
\frac{e^{-i (\mathbf{k}_{\nu} \cdot \mathbf{x}-E_{\nu } t )}}{L^{3/2}}
U_{\nu},
\end{equation}
where $U_\nu$ is the same as in Eq.~\eqref{eq:U-nu}.

\section{Cross sections and differential reaction rates
\label{sec:rates}}

As discussed in Sec.~\ref{sec:energies}, $B\sim 10^{16}\text{ G}$ will
not affect the weak interaction, which is still
described by the effective four-fermion Lagrangian
\begin{equation}
\mathcal{L}_{\text{int}} = \frac{G_{F}\cos\theta_{C}}{\sqrt{2}}
\left(N_{\alpha}^{\dagger}L^{\alpha}
 + N^{\alpha}L_{\alpha}^{\dagger}\right),
\label{eq:lagrangian}
\end{equation}
where $G_F=(292.8\,\text{GeV})^{-2}$ is the Fermi constant,
$\theta_C$ is the Cabbibo angle ($\cos^2\theta_C = 0.95$),
the leptonic charged current $L^{\alpha}$ is
\begin{equation}
L^{\alpha} = \bar{\psi}_{\nu} \gamma^{\alpha} (1-\gamma_{5}) \psi_{e},
\end{equation}
and the nucleonic current $N^{\alpha}$ is
\begin{equation}
N^{\alpha} = \bar{\psi}_{p}
\left[f \gamma^{\alpha} - g \gamma^{\alpha} \gamma_{5} +
\frac{i f_{2}}{2m_{p}} \sigma^{\alpha\beta}
\left(-i \overleftrightarrow{\mathcal{D}}_{\beta} \right) 
\right]\psi_{n}.
\label{eq:nucl-current}
\end{equation}
In the above equations, $\gamma^{\alpha}$, $\gamma_{5}$, and
$\sigma^{\alpha\beta}$ are the standard matrices describing fermionic
transitions in Dirac-Pauli
representation, and $f=1$, $g=1.26$, and $f_{2}=3.7$ are the
nucleon form factors. [A more up-to-date value of $g$ is 1.27
\cite{Eidelman:2004}. This value is recommended for calculating
the rates of the processes in Eq.~\eqref{eq:processes} for specific
application to supernova nucleosynthesis.]
The term involving $f_2$ in 
Eq.~\eqref{eq:nucl-current} represents weak magnetism and must
be included for calculations to $\mathcal{O}(1/m_N)$.
The covariant derivative 
$-i\overleftrightarrow{\mathcal{D}}_{\beta}$
in this term preserves the gauge invariance and
operates according to
\begin{equation}
\bar{\psi}_{p} O 
\left(-i \overleftrightarrow{\mathcal{D}}_{\beta} \right) \psi_{n}
= \left[\left(-i \partial_{\beta} - e A_{\beta} \right) \bar{\psi}_{p}
\right] O \psi_{n} - \bar{\psi}_{p} O 
\left(i \partial_{\beta} \psi_{n} \right),
\label{eq:covariant-d}
\end{equation}
where $O$ is a constant matrix
and $A_{\beta}$ corresponds to the electromagnetic field
($A_0=0$ here). 

Based on the above description of the weak interaction,
we derive below the cross sections of the neutrino absorption 
processes in Eq.~\eqref{eq:processes} and the differential 
reaction rates of the reverse neutrino emission processes.
We will include the magnetic field effects on particle energies and
wave functions and focus on corrections of
$\mathcal{O}(1/m_N)$ in both the transition amplitude and 
kinematics. Radiative corrections and the effect of the Coulomb field
of the proton on the electron wave function are ignored for simplicity. 
(The Coulomb field will modify the Landau wave function of the 
electron, thus making the calculation much more complicated.)
We propose an approximate treatment of these factors at the end of 
Sec.~\ref{sec:rnuabs}.

\subsection{Cross sections for neutrino absorption
\label{sec:cross}}

We first derive the cross section of $\nu_{e}+n\rightarrow e^{-}+p$
in detail. The transition matrix of this process is
\begin{equation}
\mathcal{T}_{\nu_{e}n} =  \frac{G_{F}\cos\theta_{C}}{\sqrt{2}}
\int \bar{\psi}_{p} \left[f \gamma^{\alpha} - g \gamma^{\alpha} \gamma_{5}
 + \frac{i f_{2}}{2m_{p}} \sigma^{\alpha\beta}
\left(-i \overleftrightarrow{\mathcal{D}}_{\beta}\right)
\right] \psi_{n} \, 
\bar{\psi}_{e^{-}} \gamma_{\alpha} (1 - \gamma_{5}) \psi_{\nu_{e}} \,
\ud^{4}x.
\label{eq:nun-t-matrix}
\end{equation}
With the wave functions given in Sec.~\ref{sec:energies}, 
Eq.~\eqref{eq:nun-t-matrix} can be rewritten as
\begin{equation}
\mathcal{T}_{\nu_{e}n} =
\frac{G_{F}\cos\theta_{C}}{\sqrt{2}}
\frac{eB}{2\pi L^{4}} 2\pi\delta(E_{e}+E_{p}-E_{\nu}-E_{n})
2\pi\delta(k_{ez}+k_{pz}-k_{\nu z}-k_{nz})
\mathfrak{M}_{\nu_{e}n}.
\label{eq:nun-t-matrix2}
\end{equation}
The amplitude $\mathfrak{M}_{\nu_{e}n}$ in
Eq.~\eqref{eq:nun-t-matrix2} is
\begin{align}
\mathfrak{M}_{\nu_{e}n} & = \int_{0}^{\infty} \xi\,\ud\xi
\int_{0}^{2\pi} e^{i \mathbf{w}_{\perp} \cdot \mathbf{x}_{\perp}}
e^{-i (n_{e} - r_{e} - n_{p} + r_{p}) \phi}
\Big\{ \bar{U}_{p} \gamma^{\alpha} (f - g \gamma_{5}) U_{n} \,
\bar{U}_{e^{-}} \gamma_{\alpha} (1 - \gamma_{5}) U_{\nu_{e}} \nonumber \\
 &\phantom{=} + \frac{i f_{2}}{2m_{p}}\left[ 
(X_{p})_{\beta}^{\dagger} \gamma^{0} \sigma^{\alpha\beta} U_{n}
 - (k_{n})_{\beta} \bar{U}_{p} \sigma^{\alpha\beta} U_{n}
\right] \, \bar{U}_{e^{-}} \gamma_{\alpha} (1 - \gamma_{5}) U_{\nu_e}\Big\} 
\, \ud\phi,
\label{eq:nun-amp}
\end{align}
where $\mathbf{w}=\mathbf{k}_{n}+\mathbf{k}_{\nu}$ is the
total momentum, the subscript $\perp$ denotes a vector
in the $xy$-plane, and 
\begin{equation}
(X_{p})_{\beta} \equiv 
\left[\frac{e^{i(k_{pz}z-E_{p}t)}e^{i(r_{p}-n_{p})\phi}}{%
\sqrt{2\pi L /eB}}\right]^{-1}
\left(i \partial_{\beta} - e A_{\beta} \right) \psi_{p}.
\label{eq:X-p}
\end{equation}
Evaluation of $(X_{p})_{\beta}$ for $\beta=1$ and 2 ($x$ and $y$) can be 
simplified by using Eqs.~\eqref{eq:raising-rel} and 
\eqref{eq:lowering-rel} \cite{Duan:2004th}.

The $\delta$-functions in Eq.~\eqref{eq:nun-t-matrix2} enforce conservation
of energy and of momentum in the $z$-direction, for which both the neutron 
and the proton momenta must be taken into
account in calculations to $\mathcal{O}(1/m_N)$. For 
$\nu_{e}+n\rightarrow e^{-}+p$ occurring in the material above the 
protoneutron star, the neutron momentum is especially important as
the typical value
$k_n\sim \sqrt{2 m_n T} = 43 (T / \text{MeV})^{1/2}\,\text{MeV}$
is larger than the typical $\nu_e$ momentum $k_\nu=E_\nu\sim 10$ MeV.
To account for this, we average the cross section 
over the normalized thermal distribution function
$f_{n}(\mathbf{k}_n, s_n)$ for the neutrons and obtain
\begin{equation}
\sigma_{\nu_{e}n}^{(1,B)}= 
\sum_{s_{n}}\int f_{n}(\mathbf{k}_n, s_n)\,\ud^{3}k_{n}
\sum_{s_{p},n_{p},r_{p}} \int \frac{L\ud k_{pz}}{2\pi}
\sum_{s_{e},n_{e},r_{e}} \int \frac{L\ud k_{ez}}{2\pi}
\frac{1}{L^{-3}L^{-3}} \frac{|\mathcal{T}_{\nu_{e}n}|^{2}}{\tau L^{3}},
\label{eq:sigma-nun-b-gen}
\end{equation}
where the superscript $(1,B)$ denotes the cross section
to $\mathcal{O}(1/m_N)$ and in the presence of a magnetic field,
$\tau$ is the duration of the interaction, and 
\begin{equation}
f_{n}(\mathbf{k}_{n},s_{n}) = 
\frac{e^{-(E_n-m_n)/T}}{%
\left(2\pi m_{n}T\right)^{3/2} \left(e^{\mu_{n}B/T}+e^{-\mu_{n}B/T}\right)}.
\end{equation}
The summation and integration in Eq.~\eqref{eq:sigma-nun-b-gen} must be 
treated as nested integrals. For example, the summation over $s_n$ and the
integration over $\mathbf{k}_n$ apply to not only $f_{n}(\mathbf{k}_n, s_n)$
but also the subsequent terms that have implicit dependence on $s_n$ and
$\mathbf{k}_n$. The summation and integration in the equations below should
be interpreted similarly.

Using
\begin{align}
&\phantom{=}\int\delta(E_e+E_p-E_\nu-E_n) \,
\delta(k_{ez}+k_{pz}-k_{\nu z}-k_{nz})\,\ud^3k_n\nonumber\\
&= \int\ud\Phi_n 
\int \delta(E_e+E_p-E_\nu-E_n)\,\ud\!\left(\frac{k_{n \perp}^2}{2}\right)
\int \delta(k_{ez}+k_{pz}-k_{\nu z}-k_{nz})\,\ud k_{nz} \nonumber\\
&= m_n \int\ud\Phi_n
\label{eq:rm-delta}
\end{align}
to integrate over the neutron momentum, we can rewrite
Eq.~\eqref{eq:sigma-nun-b-gen} as
\begin{align}
\sigma_{\nu_{e}n}^{(1,B)}
&= \frac{G_{F}^{2} \cos^{2}\theta_{C}}{4\pi}
  \frac{m_{n}eB}{(2\pi m_{n}T)^{3/2}}
  \frac{1}{e^{\mu_{n}B/T} + e^{-\mu_{n}B/T}}\nonumber \\
&\phantom{=}\times\sum_{n_{e}=0}^\infty
  \sum_{n_{p}=0}^\infty 
  \int_{-\infty}^{+\infty}\ud k_{ez}
  \sum_{s_{n}=\pm 1} \sum_{s_{p}=\pm 1}
  \int\limits_K \ud k_{pz}
  \int_0^{2\pi} e^{-(E_n-m_n)/T}\,\mathcal{W}_{\nu_{e}n}\,\ud\Phi_{n},
\label{eq:sigma-nun-b}
\end{align}
where
\begin{equation}
\mathcal{W}_{\nu_{e}n} \equiv \frac{eB}{2\pi L^{2}}
\sum_{s_{e},r_{e},r_{p}}|\mathfrak{M}_{\nu_{e}n}|^{2}
\label{eq:W-nun-def}
\end{equation}
is the reduced amplitude squared 
given explicitly in Appendix~\ref{sec:amp-sqr}. It follows from
Eq.~\eqref{eq:rm-delta} that $E_n$ and $k_{nz}$ in the integrand in
Eq.~\eqref{eq:sigma-nun-b} are determined in terms of other quantities
by conservation of energy and of momentum in the $z$-direction.
The integration region $K$ of $\int\ud k_{pz}$ in 
Eq.~\eqref{eq:sigma-nun-b}
is also set by these conservation laws, which require
\begin{align}
E_\nu + m_n + \frac{k_{n\perp}^2}{2 m_n} 
+ \frac{(k_{ez}+k_{pz}-k_{\nu z})^2}{2 m_n} - s_n \mu_n B
&= \sqrt{m_e^2 + k_{ez}^2 + 2 n_e e B}\nonumber \\
&\phantom{=} + m_p + \frac{k_{pz}^2}{2 m_p} + 
\frac{n_p eB}{m_p} - s_p\tilde\mu_p B.
\label{eq:conservation}
\end{align}
The above equation can be rearranged into the form
\begin{equation}
a k_{pz}^2 + b k_{pz} + c = \frac{k_{n\perp}^2}{2 m_n} \ge 0,
\end{equation}
where
\begin{subequations}
\begin{align}
a &= \frac{\Delta}{2 m_p m_n}, \\
b &= \frac{k_{\nu z}-k_{ez}}{m_n}, \\
c &= \sqrt{m_e^2 + k_{ez}^2 + 2 n_e e B}
- E_\nu - \Delta   
- \frac{(k_{\nu z}-k_{e z})^2}{2 m_n}+ \frac{n_p eB}{m_p}
- s_p \tilde\mu_p B + s_n \mu_n B,\\
\Delta&\equiv m_n-m_p.
\end{align}
\end{subequations}
Thus,
\begin{equation}
K=\begin{cases}
(-\infty, +\infty), & \text{if $b^2 \le 4ac$,} \\
\left(-\infty, (k_{pz})_-\right] \cup \left[(k_{pz})_+, +\infty\right), & 
\text{if $b^2 > 4ac$,}
\end{cases}
\end{equation}
where
\begin{equation}
(k_{pz})_\pm = \frac{-b \pm \sqrt{b^2 - 4ac}}{2a}.
\label{eq:kpz-limits}
\end{equation}

For $\bar{\nu}_{e}+p\rightarrow e^{+}+n$
occurring in the material above the protoneutron star,
the cross section can be written as
\begin{equation}
\sigma_{\bar{\nu}_{e}p}^{(1,B)} = 
\overline{\sum_{r_{p}}}\sum_{s_{p},n_{p}} 
\int f_{p}(k_{pz},n_{p},s_{p})\,\ud k_{pz}
\sum_{s_{n}}\int\frac{L^{3}\ud^{3}k_{n}}{(2\pi)^{3}}
\sum_{s_{e},n_{e},r_{e}}\int\frac{L\ud k_{ez}}{2\pi}
\frac{1}{L^{-3}L^{-3}}
\frac{|\mathcal{T}_{\bar\nu_{e}p}|^{2}}{\tau L^{3}},
\end{equation}
where \begin{equation}
\overline{\sum_{r_{p}}} \equiv
\left(\frac{eB L^{2}}{2\pi}\right)^{-1}
\sum_{r_{p}}
\label{eq:avg-rp}
\end{equation}
and 
\begin{equation}
f_{p}(k_{pz},n_{p},s_{p}) = 
\frac{e^{-(E_p-m_p)/T}}{\sqrt{2\pi m_{p}T}}
\frac{1-e^{-eB/m_{p}T}}{%
e^{\tilde{\mu}_{p}B/T} + e^{-(\tilde{\mu}_{p}B/T) - (eB/m_{p}T)}}
\label{eq:proton-distribution}
\end{equation}
is the normalized thermal distribution function for the protons.
Using again the integration over the neutron momentum to get rid of the 
$\delta$-functions, we obtain
\begin{align}
\sigma_{\bar{\nu}_{e}p}^{(1,B)} & =
\frac{G_{F}^{2}\cos^{2}\theta_{C}}{8\pi^{2}}
\frac{m_{n}}{\sqrt{2\pi m_p T}}
\frac{1-e^{-eB/m_{p}T}}{%
e^{\tilde{\mu}_{p}B/T} + e^{-(\tilde{\mu}_{p}B/T) - (eB/m_{p}T)}}\nonumber \\
& \phantom{=}\times\sum_{n_{e}=0}^{\infty}
\sum_{n_{p}=0}^{\infty} \int_{-\infty}^{+\infty}\ud k_{ez}
\sum_{s_{n}=\pm 1}\sum_{s_{p}=\pm 1}
\int\limits_{K^\prime} \ud k_{pz}
\int_0^{2 \pi} e^{-(E_p-m_p)/T}\,\mathcal{W}_{\bar{\nu}_{e}p}\,\ud\Phi_{n}.
\label{eq:sigma-nup-b}
\end{align}
The reduced amplitude squared $\mathcal{W}_{\bar{\nu}_{e}p}$ in the above
equation can be obtained from 
$\mathcal{W}_{\nu_{e}n}$ by making the substitution
\begin{equation}
\begin{split}
(E_{\nu},\mathbf{k}_{\nu}) & \longrightarrow (-E_{\nu},-\mathbf{k}_{\nu}),\\
(E_{e},k_{ez}) & \longrightarrow (-E_{e},-k_{ez}).\\
\end{split}
\label{eq:xing-rel}
\end{equation}
The integration region $K^\prime$ of $\int\ud k_{pz}$ in 
Eq.~\eqref{eq:sigma-nup-b} is determined from energy and momentum 
conservation as in Eq.~\eqref{eq:conservation} but with the above 
substitution implemented.

For application to supernova neutrinos, it is useful to further average
the cross sections in Eqs.~\eqref{eq:sigma-nun-b} and \eqref{eq:sigma-nup-b}
over the relevant normalized neutrino energy spectra $f_{\nu}(E_{\nu})$
to obtain
\begin{equation}
\langle\sigma_{\nu N}\rangle =
\int \sigma_{\nu N} f_\nu(E_\nu)\,\ud E_\nu,
\label{eq:sigma-nuN-avg}
\end{equation}
where $\sigma_{\nu N}$ stands for $\sigma_{{\nu}_{e}n}^{(1,B)}$ or
$\sigma_{\bar{\nu}_{e}p}^{(1,B)}$.
A typical form of $f_\nu(E_\nu)$ adopted in the literature is
\begin{equation}
f_{\nu}(E_{\nu}) = \frac{1}{T_\nu^3F_2(\eta_\nu)} \frac{E_{\nu}^{2}}{%
\exp\left[\left(E_{\nu}/T_{\nu}\right)-\eta_{\nu}\right]+1},
\label{eq:nuspec}
\end{equation}
where $T_{\nu}$ and $\eta_{\nu}$ are constant parameters and
\begin{equation}
F_n(\eta_\nu)\equiv \int_0^\infty \frac{x^{n}}{%
\exp\left(x-\eta_{\nu}\right)+1}\,\ud x.
\end{equation}
For the neutrino energy spectra in Eq.~\eqref{eq:nuspec}, 
the average neutrino energy is
\begin{equation}
\langle E_\nu\rangle=\frac{F_3(\eta_\nu)}{F_2(\eta_\nu)}T_{\nu}.
\end{equation}

\subsection{Differential reaction rates for neutrino emission
\label{sec:diff}}

As can be seen from Eqs.~\eqref{eq:sigma-nun-b}, \eqref{eq:sigma-nup-b},
and \eqref{eq:sigma-nuN-avg}, the cross sections
$\langle \sigma_{\nu_{e}n}^{(1,B)}\rangle $
and $\langle \sigma_{\bar{\nu}_{e}p}^{(1,B)}\rangle $
for the neutrino absorption processes $\nu_{e}+n\rightarrow e^{-}+p$
and $\bar{\nu}_{e}+p\rightarrow e^{+}+n$, respectively, have the same
generic form
\begin{equation}
\int\ud E_{\nu}\sum_{n_{e}=0}^{\infty}
\sum_{n_{p}=0}^{\infty}\int_{-\infty}^{+\infty}\ud k_{ez}
\sum_{s_{n}=\pm1}\sum_{s_{p}=\pm1}\int\limits _{\tilde{K}}\ud k_{pz}
\int_{0}^{2\pi}\mathcal{F}\,\ud\Phi_{n},
\label{eq:generic}
\end{equation}
where $\tilde{K}=K\text{ or }K^{\prime}$, and $\mathcal{F}$ is the
integrand involving the relevant amplitude squared and distribution 
functions. If we use the differential reaction rates with respect to 
$\cos\Theta_\nu$ to describe the neutrino emission processes 
$e^- + p \rightarrow \nu_e + n$ and $e^{+}+n\rightarrow\bar{\nu}_{e}+p$,
then these rates also have the generic form in Eq.~\eqref{eq:generic}.
This follows from the symmetry between the forward and reverse processes.
In particular,  the transition amplitudes
squared $\left|\mathcal{T}_{e^-p}\right|^{2}$ and
$\left|\mathcal{T}_{e^+n}\right|^{2}$ are identical to
$\left|\mathcal{T}_{\nu_{e}n}\right|^{2}$ and
$\left|\mathcal{T}_{\bar\nu_{e}p}\right|^{2}$, respectively.
By taking advantage of the symmetry between the neutrino absorption
and emission processes, numerical computation of the cross sections
for the former and the differential reaction rates for the latter is
greatly simplified.

For $e^- + p \rightarrow \nu_e + n$ occurring in the material above 
the protoneutron star, the differential reaction rate is
\begin{eqnarray}
\frac{\ud\lambda_{e^{-}p}^{(1,B)}}{\ud\cos\Theta_{\nu}} & = 
& \overline{\sum_{r_{p}}}\sum_{s_{p},n_{p}}
\int f_{p}(k_{pz},n_{p},s_{p})\,\ud k_{pz}
\sum_{s_{e},n_{e},r_{e}}
\int\frac{L\ud k_{ez}}{2\pi}
\frac{1}{L^{3}}\frac{1}{e^{(E_{e}/T)-\eta_{e}}+1}
\nonumber \\
 &  & \times\int\frac{L^{3}E_{\nu}^{2}\ud E_{\nu}}{4\pi^{2}}
\sum_{s_{n}}\int\frac{L^{3}\ud^{3}k_{n}}{(2\pi)^{3}}
\frac{1}{L^{-3}L^{-3}}
\frac{\left|\mathcal{T}_{\nu_{e}n}\right|^{2}}{\tau L^{3}},
\label{eq:diff-lambda-ep-gen}
\end{eqnarray}
where $\eta_{e}$ is the degeneracy parameter characterizing
the Fermi-Dirac distribution function of the electrons.
Integrating over the neutron momentum as in 
Eq.~\eqref{eq:rm-delta}, we obtain
\begin{eqnarray}
\frac{\ud\lambda_{e^{-}p}^{(1,B)}}{\ud\cos\Theta_{\nu}} & = & 
\frac{G_{F}^{2}\cos^{2}\theta_{C}}{32\pi^{4}}
\frac{m_{n}}{\sqrt{2\pi m_{p}T}}
\frac{1-e^{-eB/m_{p}T}}{%
e^{\tilde{\mu}_{p}BT}+e^{-(\tilde{\mu}_{p}B/T)-(eB/m_{p}T)}}
\int_{0}^{\infty}E_{\nu}^{2}\ud E_{\nu}
\nonumber \\
 &  & \times\sum_{n_{e}=0}^{\infty}
\sum_{n_{p}=0}^{\infty}
\int_{-\infty}^{+\infty}\ud k_{ez}
\sum_{s_{n}=\pm1}\sum_{s_{p}=\pm1}\int\limits _{K}\ud k_{pz}
\nonumber \\
 &  & \times
\int_{0}^{2\pi}\frac{e^{-(E_p-m_p)/T}}{e^{(E_{e}/T)-\eta_{e}}+1}\,
\mathcal{W}_{\nu_{e}n}\,\ud\Phi_{n}.
\end{eqnarray}
Similarly, we obtain the differential reaction rate for 
$e^{+}+n\rightarrow\bar{\nu}_{e}+p$ as
\begin{eqnarray}
\frac{\ud\lambda_{e^{+}n}^{(1,B)}}{\ud\cos\Theta_{\nu}} & = & 
\sum_{s_{n}}\int f_{n}(\mathbf{k}_{n},s_{n})\,\ud^{3}k_{n}
\sum_{s_{e},n_{e},r_{e}}
\int\frac{L\ud k_{ez}}{2\pi}
\frac{1}{L^{3}}\frac{1}{e^{(E_{e}/T)+\eta_{e}}+1}
\nonumber \\
 &  & \times\int\frac{L^{3}E_{\nu}^{2}\ud E_{\nu}}{4\pi^{2}}
\sum_{s_{p},n_{p},r_{p}}\int\frac{L\ud k_{pz}}{2\pi}
\frac{1}{L^{-3}L^{-3}}
\frac{\left|\mathcal{T}_{\bar{\nu}_{e}p}\right|^{2}}{\tau L^{3}}
\nonumber \\
 & = & \frac{G_{F}^{2}\cos^{2}\theta_{C}}{16\pi^{3}}
\frac{m_{n}eB}{(2\pi m_{n}T)^{3/2}}
\frac{1}{e^{\mu_{n}B/T}+e^{-\mu_{n}B/T}}
\int_{0}^{\infty}E_{\nu}^{2}\,\ud E_{\nu}
\nonumber \\
 &  & \times\sum_{n_{e}=0}^{\infty}\sum_{n_{p}=0}^{\infty}
\int_{-\infty}^{+\infty}\ud k_{ez}
\sum_{s_{n}=\pm1}\sum_{s_{p}=\pm1}\int\limits _{K^{\prime}}\ud k_{pz}
\nonumber \\
 &  & \times
\int_{0}^{2\pi}\frac{e^{-(E_n-m_n)/T}}{e^{(E_{e}/T)+\eta_{e}}+1}\,
\mathcal{W}_{\bar{\nu}_{e}p}\,\ud\Phi_{n}.
\end{eqnarray}

\section{Rates of Neutrino Processes in Supernovae\label{sec:results}}

We now calculate the rates of the neutrino absorption and emission
processes in Eq.~\eqref{eq:processes} for the supernova environment
near a protoneutron star that possesses a strong magnetic field.
A wide range of heavy elements may be produced during the ejection
of the material above the protoneutron star. As mentioned in the
introduction, a key parameter governing this production is the 
neutron-to-proton ratio of the
material \cite{Woosley:1992,Hoffman:1996}, which depends on the 
competition between the neutrino absorption and emission
processes \cite{Qian:1996}. We will calculate the rates of these
processes in the context of heavy element nucleosynthesis, for which
the accuracy of these rates is especially important. In this context,
the material above the protoneutron star is characterized by
temperatures of $T\sim 1$~MeV, entropies of $S\sim 100$ (in units of
Boltzmann constant per nucleon), and electron fractions of
$Y_e\lesssim 0.5$. For these conditions, the nucleons in the material
are nonrelativistic and nondegenerate while the $e^\pm$ are 
relativistic and have a small degeneracy parameter $0<\eta_e\ll 1$.
The thermal distribution functions of the nucleons and $e^\pm$ have 
been given in Sec.~\ref{sec:rates}. The neutrinos emitted from
the protoneutron star are not in thermal equilibrium with the
overlying material and their energy distribution functions are taken 
to be of the form in Eq.~\eqref{eq:nuspec}. As discussed in 
Ref.~\cite{Duan:2004nc}, Pauli blocking for the final states of the 
neutrino processes above the protoneutron star is unimportant
and will be ignored.

\subsection{Neutrino absorption\label{sec:rnuabs}}

At a radius $R$ above the protoneutron star, the rate of neutrino
absorption per nucleon can be estimated as
\begin{equation}
\lambda_{\nu N}=\frac{L_\nu\langle\sigma_{\nu N}\rangle}
{4\pi R^2\langle E_\nu\rangle}=49.7
\left(\frac{L_\nu}{10^{51}\ {\rm erg\ s}^{-1}}\right)
\left(\frac{10\ {\rm MeV}}{\langle E_\nu\rangle}\right)
\left(\frac{\langle\sigma_{\nu N}\rangle}{10^{-41}\ {\rm cm}^2}\right)
\left(\frac{10\ {\rm km}}{R}\right)^2\ {\rm s}^{-1},
\end{equation}
where $L_\nu$ is the neutrino luminosity and has a typical value of
$\sim 10^{51}$ erg s$^{-1}$ in the supernova epoch of interest. The key 
quantity $\langle\sigma_{\nu N}\rangle$ in the above equation is 
obtained by averaging $\sigma_{\nu N}$ over the neutrino energy spectrum.
We first compare various approximations for $\sigma_{\nu N}$ as functions 
of the neutrino energy $E_\nu$.

The cross sections for neutrino absorption on nucleons in a magnetic 
field have been derived to $\mathcal{O}(1/m_N)$ as 
$\sigma_{\nu N}^{(1,B)}$ in Sec.~\ref{sec:cross}.
To $\mathcal{O}(1)$, the zeroth order in $1/m_N$, the cross sections 
are \cite{Duan:2004nc}
\begin{align}
\sigma_{\nu N}^{(0,B)} & =
\sigma_{B,1} 
\left[ 1 + 2 \chi \frac{(f \pm g) g}{f^{2} + 3 g^{2}} 
\cos \Theta_{\nu } \right] \nonumber \\
 &\phantom{=} + \sigma_{B,2} 
\left[ \frac{f^{2} - g^{2}}{f^{2} + 3 g^{2}}
\cos \Theta_{\nu} + 2 \chi \frac{(f \mp g)}{f^{2} + 3 g^{2}} \right],
\label{eq:sigma0-b}
\end{align}
where
\begin{align}
\sigma_{B,1} & = 
\frac{G_{F}^{2} \cos^{2} \theta_{C}}{2 \pi}(f^{2} + 3 g^{2}) eB
\sum_{n_e=0}^{n_{e,\text{max}}} 
\frac{g_{n_e} E_{e}^{(0)}}{\sqrt{(E_{e}^{(0)})^{2} - m_{e}^{2} - 2n_eeB}}
\label{eq:sigb1},\\
\sigma_{B,2} & = 
\frac{G_{F}^{2} \cos^{2} \theta_{C}}{2 \pi}(f^{2} + 3 g^{2}) eB
\frac{E_{e}^{(0)}}{\sqrt{(E_{e}^{(0)})^{2} - m_{e}^{2}}},
\label{eq:sigb2}\\
E_{e}^{(0)}&=E_{\nu}\pm\Delta,\\
n_{e,\text{max}} &= \left[\frac{(E_e^{(0)})^2-m_e^2}{2eB}\right]_\text{int}.
\end{align}
In the above equations and elsewhere in this subsection,
the upper sign is for $\nu_{e}+n\rightarrow e^{-}+p$ and
the lower sign is for $\bar{\nu}_{e}+p\rightarrow e^{+}+n$.
In Eq.~\eqref{eq:sigma0-b},
\begin{equation}
\chi=\frac{\exp(\mu B/T)-\exp(-\mu B/T)}{\exp(\mu B/T)+\exp(-\mu B/T)}
\label{eq:chi}
\end{equation}
is the polarization of nucleon spin, where $\mu$ is the magnetic
moment of the relevant nucleon with $\mu_p=2.79\mu_N$ and
$\mu_n=-1.91\mu_N$. For the case of interest here,
$|\mu|B/T\ll 1$, so
\begin{equation}
\chi=\frac{\mu B}{T}=3.15\times 10^{-2}\left(\frac{\mu}{\mu_N}\right)
\left(\frac{B}{10^{16}\ \mathrm{G}}\right)\left(\frac{\text{MeV}}{T}\right).
\end{equation}

The term proportional to $\sigma_{B,2}$ in Eq.~\eqref{eq:sigma0-b}
arises because the ground Landau level of $e^\pm$ has only one spin 
state while any other level has two. In Eq.~\eqref{eq:sigb1} for
$\sigma_{B,1}$, the product of $eB$ and the sum gives the total phase 
space of the $e^\pm$ in the final state. In the limit
$n_{e,\text{max}}\gg 1$, the summation of the Landau levels can be 
replaced by integration and $\sigma_{B,1}$ approaches  
\begin{equation}
\sigma_{\nu N}^{(0)} = \frac{G_{F}^{2}\cos^{2}\theta_{C}}{\pi}
E_{e}^{(0)}\sqrt{(E_{e}^{(0)})^2-m_{e}^2},
\label{eq:sigma0}
\end{equation}
which is the cross section to $\mathcal{O}(1)$ in the absence of any 
magnetic field. In the same limit, $\sigma_{B,2}$ is negligible compared 
with $\sigma_{B,1}$, so $\sigma_{\nu N}^{(0,B)}$ approaches
$\sigma_{\nu N}^{(0)}(1+\epsilon_\chi)$, where
\begin{equation}
\epsilon_\chi = \chi \frac{2(f\pm g)g}{f^2 + 3g^2} \cos\Theta_\nu
\end{equation}
results from the polarization of the initial nucleon spin by the 
magnetic field.

\begin{figure*}
\includegraphics*[width=1.0\textwidth, keepaspectratio]{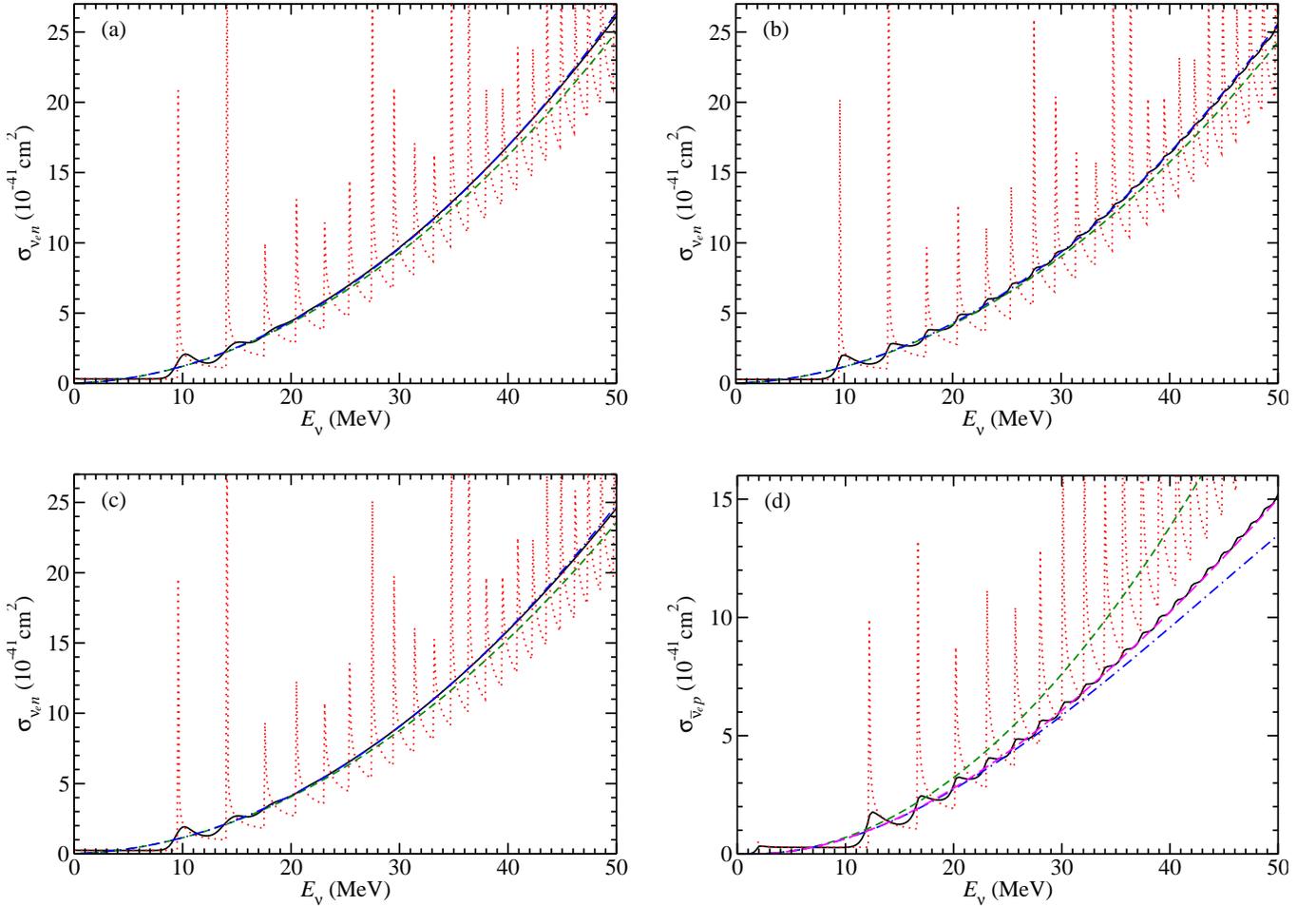}
\caption{\label{fig:nuN}%
(Color online) Comparison of various approximations for 
$\sigma_{\nu N}$ [solid curves: $\sigma^{(1,B)}_{\nu N}$; 
dotted curves: $\sigma^{(0,B)}_{\nu N}$; 
short-dashed curves: $\sigma^{(0)}_{\nu N}(1+\epsilon_{\chi})$; 
dot-dashed curves: $\sigma^{(1)}_{\nu N}(1+\epsilon_{\chi})$;
long-dashed curves: $\sigma^{(1*)}_{\nu N}(1+\epsilon_{\chi})$].
The dot-dashed and long-dashed curves for $\sigma_{\nu_en}$
are indistinguishable. The results for $\sigma_{\nu_en}$ are
shown in (a)--(c) for $\cos\Theta_{\nu}=-1$, 0, and 1, respectively,
while those for $\sigma_{\bar\nu_ep}$ are shown in (d) for
$\cos\Theta_{\nu}=0$ ($\sigma_{\bar\nu_ep}$ has little angular 
dependence). These results are calculated using 
$B=10^{16}\,\text{G}$, $T=2\,\text{MeV}$ for the matter temperature, 
$\chi_n=-0.03$, and $\chi_p=0.04$.}
\end{figure*}

For numerical examples of the cross sections, we take 
$B=10^{16}\,\text{G}$.
The cross sections $\sigma_{\nu_en}^{(0,B)}$ for 
$\cos\Theta_{\nu}=-1$, 0, and 1 as functions of $E_\nu$
are shown as the dotted curves in Figs.~\ref{fig:nuN}a, 
\ref{fig:nuN}b, and \ref{fig:nuN}c, 
respectively. The angle-dependent terms in
Eq.~\eqref{eq:sigma0-b} for $\sigma_{\bar\nu_ep}^{(0,B)}$
are proportional to the difference between $f$ and $g$.
As the numerical values of $f$ and $g$ are close, these terms
are very small. So we only show the cross section 
$\sigma_{\bar\nu_ep}^{(0,B)}$ for $\cos\Theta_{\nu}=0$
as the dotted curve in Fig.~\ref{fig:nuN}d. 
All the dotted curves in Fig.~\ref{fig:nuN} have spikes
superposed on a general trend. The varying heights of these spikes
are artifacts of the plotting tool: all the spikes should have
been infinitely high as they correspond to ``resonances'' at
$E_e^{(0)}=\sqrt{m_{e}^{2}+2n_eeB}$, for which a new Landau level
opens up. These singularities are integrable and
do not give infinite probabilities of interaction in practice.
For example, at a given $E_\nu$, the thermal motion of the absorbing 
nucleons will produce a range of $E_e$ and the cross section obtained
from integration over this range will be finite. Thus,
the spikes in $\sigma_{\nu N}^{(0,B)}$ will be smeared out by
the thermal motion of the absorbing nucleons, which is
similar to the Doppler broadening of the photon absorption
lines in the solar light spectrum. The effects of such motion 
are of $\mathcal{O}(1/m_N)$ and have been taken into account 
by the cross sections $\sigma_{\nu N}^{(1,B)}$ derived in 
Sec.~\ref{sec:cross}. Using $T=2$ MeV for illustration, we show
$\sigma_{\nu N}^{(1,B)}$ as the solid curves in Fig.~\ref{fig:nuN}. 
It can be
seen that where spikes occur in $\sigma_{\nu N}^{(0,B)}$, there
are only smooth bumps in $\sigma_{\nu N}^{(1,B)}$. Clearly,
$\sigma_{\nu N}^{(1,B)}$ is more physical than
$\sigma_{\nu N}^{(0,B)}$.

Two more aspects of Fig.~\ref{fig:nuN} require discussion. First, 
the bumps in $\sigma_{\nu N}^{(1,B)}$ diminish as $E_\nu$ increases 
and become invisible for $E_\nu\gg \sqrt{eB}\sim 8$ MeV. This is
expected from the correspondence principle: when a number of
Landau levels for $e^\pm$ and protons can be occupied, the
quantum effects of the magnetic field are small. As noted in
Sec.~\ref{sec:pwave}, the absorbing proton in 
$\bar\nu_e + p \rightarrow e^+ + n$ can occupy many levels 
for $T\sim 1$~MeV. However, for the $e^+$ in 
$\bar\nu_e + p \rightarrow e^+ + n$ and the $e^-$ and the proton
in $\nu_e + n \rightarrow e^- + p$, occupation of many levels
requires $E_\nu\gg \sqrt{eB}\sim 8$ MeV (see Secs.~\ref{sec:ewave}
and \ref{sec:pwave}). Second, while the general trends
of the dotted curves for $\sigma_{\nu_en}^{(0,B)}$ appear to
follow the corresponding solid curves for $\sigma_{\nu_en}^{(1,B)}$,
the general trend of the dotted curve for 
$\sigma_{\bar\nu_ep}^{(0,B)}$ deviates substantially from the
corresponding solid curve for $\sigma_{\bar\nu_ep}^{(1,B)}$. 
This concerns the effects of weak magnetism and recoil of the 
final-state nucleons, both of which are of $\mathcal{O}(1/m_N)$ and
are taken into account by $\sigma_{\nu N}^{(1,B)}$ but not by
$\sigma_{\nu N}^{(0,B)}$. Figure~\ref{fig:nuN} shows that these effects 
give small corrections to $\sigma_{\nu_en}^{(0,B)}$ but
much larger corrections to $\sigma_{\bar\nu_ep}^{(0,B)}$.

To better understand the effects of weak magnetism and recoil of 
the final-state nucleons, we make use of the correspondence
principle. As noted above, the effects of Landau levels become
negligible for $E_\nu\gg \sqrt{eB}\sim 8$ MeV. In this case,
the only surviving quantum effect of the magnetic 
field is polarization of the initial nucleon spin, which gives
rise to a dependence on $\cos\Theta_\nu$ for the cross sections 
due to parity violation of weak interaction. Thus, allowing for
this surviving effect, we should recover the results for
no magnetic field in the limit of high $E_\nu$. In the absence
of any field, the cross sections 
$\sigma_{\nu N}^{(1)}$ to $\mathcal{O}(1/m_N)$ is
(see, e.g., Refs.~\cite{Vogel:1999zy,Horowitz:2001xf})
\begin{equation}
\sigma_{\nu N}^{(1)} = \sigma_{\nu N}^{(0)}
\left\{1-\frac{2[f^2 \mp 2(f+f_2)g + 5g^2]}{f^2+3g^2}\frac{E_\nu}{m_N}
\right\},\label{eq:sigma-nuN-1}
\end{equation}
where we have ignored terms like $m_e^2/E_\nu^2$ and $\Delta/m_N$.
The above zero-field cross sections assume that the initial
nucleon spin is unpolarized, and therefore, do not depend on
$\cos\Theta_\nu$. If a small polarization $\chi$ is artificially
imposed, the modified cross sections should have an additional factor 
$1+\epsilon_\chi$.
The term proportional to $E_\nu/m_N$ in Eq.~\eqref{eq:sigma-nuN-1}
represents the effects of weak magnetism and recoil of 
the final-state nucleons. The coefficient in this term is
1.01 for $\nu_e + n \rightarrow e^- + p$ and $-7.21$ for
$\bar\nu_e + p \rightarrow e^+ + n$. Therefore, over the range
$E_\nu\sim 10$--50 MeV typical of supernova neutrinos, the
correction due to the above effects is $\sim 1$--5\% for the
former reaction but amounts to $\sim -7$\% to $-36$\% for the 
latter reaction. The importance of these corrections has been
discussed in other contexts \cite{Horowitz:1999,Vogel:1999zy}.

Using $\chi_n=-0.03$ and $\chi_p=0.04$ corresponding to
$B=10^{16}$ G and $T=2$ MeV, we show 
$\sigma_{\nu N}^{(0)}(1+\epsilon_\chi)$ and
$\sigma_{\nu N}^{(1)}(1+\epsilon_\chi)$ as the short-dashed and
dot-dashed curves, respectively, in Fig.~\ref{fig:nuN}. The small
increase from $\sigma_{\nu_en}^{(0)}$ to $\sigma_{\nu_en}^{(1)}$
and the much larger decrease from $\sigma_{\bar\nu_ep}^{(0)}$ 
to $\sigma_{\bar\nu_ep}^{(1)}$ given in Eq.~\eqref{eq:sigma-nuN-1}
can be seen from this figure. In addition, as expected from 
the correspondence principle, the general trends of the dotted
curves for $\sigma_{\nu_en}^{(0,B)}$ closely follow the 
short-dashed curves for 
$\sigma_{\nu_en}^{(0)}(1+\epsilon_{\chi_n})$ 
at $E_\nu\gtrsim 20$ MeV
and the solid curves for $\sigma_{\nu_en}^{(1,B)}$ become
indistinguishable from the dot-dashed curves for
$\sigma_{\nu_en}^{(1)}(1+\epsilon_{\chi_n})$ in the same regime
(see Figs.~\ref{fig:nuN}a--c).
However, while the relation between the dotted curve for
$\sigma_{\bar\nu_ep}^{(0,B)}$ and the short-dashed curve for
$\sigma_{\bar\nu_ep}^{(0)}(1+\epsilon_{\chi_p})$ is in accordance
with the correspondence principle, the solid curve for
$\sigma_{\bar\nu_ep}^{(1,B)}$ clearly stays above the dot-dashed
curve for $\sigma_{\bar\nu_ep}^{(1)}(1+\epsilon_{\chi_p})$ at
$E_\nu\gtrsim 25$ MeV (see Fig.~\ref{fig:nuN}d). This apparent
violation of the correspondence principle for 
$\sigma_{\bar\nu_ep}^{(1,B)}$ and
$\sigma_{\bar\nu_ep}^{(1)}(1+\epsilon_{\chi_p})$ is caused by the
slightly different treatments of the reaction kinematics
in calculating $\sigma_{\bar\nu_ep}^{(1,B)}$ and 
$\sigma_{\bar\nu_ep}^{(1)}$. 

We have used the transition
amplitudes to $\mathcal{O}(1/m_N)$ in calculating both
$\sigma_{\nu N}^{(1,B)}$ and $\sigma_{\nu N}^{(1)}$.
However, we have treated the reaction kinematics exactly 
for $\sigma_{\nu N}^{(1,B)}$ [assuming nonrelativistic
nucleons, see Eqs.~(\ref{eq:conservation}--\ref{eq:kpz-limits})]
but only to $\mathcal{O}(1/m_N)$ for
$\sigma_{\nu N}^{(1)}$. This difference does not 
affect the comparison between $\sigma_{\nu_en}^{(1,B)}$ and
$\sigma_{\nu_en}^{(1)}(1+\epsilon_{\chi_n})$ as the total
correction from weak magnetism and nucleon recoil is small
to $\mathcal{O}(1/m_N)$ in this case, and the terms of orders
higher than $\mathcal{O}(1/m_N)$ are even smaller.
In contrast, the importance of the weak magnetism and recoil effects for
$\bar\nu_e + p \rightarrow e^+ + n$ enables terms of orders
higher than $\mathcal{O}(1/m_N)$ to give rather large
corrections to the cross section. Such terms are included
in $\sigma_{\bar\nu_ep}^{(1,B)}$ due to exact treatment of
the reaction kinematics but not in $\sigma_{\bar\nu_ep}^{(1)}$.
In Ref.~\cite{Strumia:2003}, the zero-field cross sections for
neutrino absorption on nucleons were derived to
$\mathcal{O}(1/m_N)$ but with reaction kinematics treated
exactly. We denote these cross sections as 
$\sigma_{\nu N}^{(1*)}$. For consistency with the rest of the
paper, we ignore radiative corrections and the effect of the
Coulomb interaction between the final-state particles for
$\nu_e + n \rightarrow e^- + p$, both of
which were taken into account
in Ref.~\cite{Strumia:2003}. It was shown in this reference
that $\sigma_{\bar{\nu}_e p}^{(1*)}$ is more accurate than
$\sigma_{\bar{\nu}_e p}^{(1)}$. We show 
$\sigma_{\nu N}^{(1*)}(1+\epsilon_{\chi})$ as the long-dashed 
curves in Fig.~\ref{fig:nuN}. It can be seen that the long-dashed
curves for $\sigma_{\nu_en}^{(1*)}(1+\epsilon_{\chi_n})$ are
indistinguishable from the corresponding dot-dashed curves for
$\sigma_{\nu_en}^{(1)}(1+\epsilon_{\chi_n})$ over the range of
$E_\nu$ shown but the long-dashed
curve for $\sigma_{\bar\nu_ep}^{(1*)}(1+\epsilon_{\chi_p})$ lies
significantly above the dot-dashed curve for
$\sigma_{\bar\nu_ep}^{(1)}(1+\epsilon_{\chi_p})$
at $E_\nu\gtrsim 25$ MeV.
In addition, the solid curves for $\sigma_{\nu_en}^{(1,B)}$
and $\sigma_{\bar{\nu}_e p}^{(1,B)}$ settle down to the 
corresponding long-dashed curves for 
$\sigma_{\nu_en}^{(1*)}(1+\epsilon_{\chi_n})$ and
$\sigma_{\bar\nu_ep}^{(1*)}(1+\epsilon_{\chi_p})$
at $E_\nu\gtrsim 20$ and 25 MeV, respectively. Thus, the
cross sections $\sigma_{\nu_en}^{(1,B)}$
and $\sigma_{\bar{\nu}_e p}^{(1,B)}$ calculated above are
in full agreement with the correspondence principle.

\begin{table*}
\caption{\label{tab:sigma-nuN-avg}%
Comparison of various approximations for 
$\langle\sigma_{\nu N}\rangle$ (in units of
$10^{-41}\,\text{cm}^{-2}$). These results are calculated using
$\langle E_{\nu_{e}}\rangle =11\,\text{MeV}$,
$\langle E_{\bar{\nu}_{e}}\rangle =16\,\text{MeV}$,
$B=10^{16}\,\text{G}$, $T=2\,\text{MeV}$ for the matter temperature, 
$\chi_n = -0.03$, and $\chi_p = 0.04$.}

\begin{ruledtabular}
\begin{tabular}{cdddddd}
& \multicolumn{3}{c}{$\nu_e+n\rightarrow e^- + p$} & 
\multicolumn{3}{c}{$\bar\nu_e+p\rightarrow e^+ + n$} \\ 
\cline{2-4} \cline{5-7}
$\cos\Theta_\nu$ & \multicolumn{1}{c}{$-1$} & \multicolumn{1}{c}{0} &
\multicolumn{1}{c}{1} & \multicolumn{1}{c}{$-1$} &
\multicolumn{1}{c}{0} & \multicolumn{1}{c}{1} \\ \hline
$\langle\sigma^{(0)}_{\nu N}\rangle(1+\epsilon_{\chi})$ 
& 1.67 & 1.62 & 1.57 & 2.49 & 2.48 & 2.46 \\
$\langle\sigma^{(1)}_{\nu N}\rangle(1+\epsilon_{\chi})$
& 1.70 & 1.65 & 1.60 & 2.06 & 2.04 & 2.02\\
$\langle\sigma^{(1*)}_{\nu N}\rangle(1+\epsilon_{\chi})$
& 1.69 & 1.65 & 1.60 & 2.11 & 2.09 & 2.07\\
$\langle\sigma^{(0,B)}_{\nu N}\rangle$ 
& 1.65 & 1.57 & 1.50 & 2.50 & 2.46 & 2.42\\
$\langle\sigma^{(1,B)}_{\nu N}\rangle$
 & 1.68 & 1.61 & 1.54 & 2.11 & 2.09 & 2.07\\
\end{tabular}
\end{ruledtabular}
\end{table*}

We now calculate the average cross sections 
$\langle\sigma_{\nu N}\rangle$ using the neutrino
energy spectra in Eq.~\eqref{eq:nuspec}.
We take $\eta_{\nu_{e}}=\eta_{\bar{\nu}_{e}}=3$,
$\langle E_{\nu_{e}}\rangle =11\,\text{MeV}$, and 
$\langle E_{\bar{\nu}_{e}}\rangle =16\,\text{MeV}$.
For these parameters, $T_{\nu_{e}}=2.75\,\text{MeV}$ and 
$T_{\bar{\nu}_{e}}=4\,\text{MeV}$. Adopting the same $B$, $T$,
$\chi_n$, and $\chi_p$ as for Fig.~\ref{fig:nuN}, we give
$\langle\sigma^{(0)}_{\nu N}\rangle(1+\epsilon_\chi)$,
$\langle\sigma^{(1)}_{\nu N}\rangle(1+\epsilon_\chi)$,
$\langle\sigma^{(1*)}_{\nu N}\rangle(1+\epsilon_\chi)$,
$\langle\sigma^{(0,B)}_{\nu N}\rangle$, and
$\langle\sigma^{(1,B)}_{\nu N}\rangle$ for
$\cos\Theta_\nu=-1$, 0, and 1, respectively, in 
Table~\ref{tab:sigma-nuN-avg}. As discussed above, 
$\sigma^{(1)}_{\nu N}$ and $\sigma^{(1*)}_{\nu N}$ differ from
$\sigma^{(0)}_{\nu N}$ due to the effects of weak magnetism 
and recoil of the final-state nucleons. These effects slightly
increase the cross sections for $\nu_e+n\rightarrow e^- + p$
but substantially decrease those for
$\bar\nu_e+p\rightarrow e^+ + n$. As can be seen from 
Table~\ref{tab:sigma-nuN-avg},
$\langle\sigma_{\nu_en}^{(1)}\rangle(1+\epsilon_{\chi_n})$ 
and
$\langle\sigma_{\nu_en}^{(1*)}\rangle(1+\epsilon_{\chi_n})$ 
are only a few percent larger than 
$\langle\sigma_{\nu_en}^{(0)}\rangle(1+\epsilon_{\chi_n})$
but 
$\langle\sigma_{\bar\nu_ep}^{(1)}\rangle(1+\epsilon_{\chi_p})$ 
and
$\langle\sigma_{\bar\nu_ep}^{(1*)}\rangle(1+\epsilon_{\chi_p})$ 
are $\sim 20$\% smaller than 
$\langle\sigma_{\bar\nu_ep}^{(0)}\rangle(1+\epsilon_{\chi_p})$.
The differences between
$\langle\sigma^{(1,B)}_{\nu N}\rangle$ and
$\langle\sigma^{(0,B)}_{\nu N}\rangle$ are similar.
On the other hand, the effects of the magnetic field on the
average cross sections are small for both
$\nu_e+n\rightarrow e^- + p$ and $\bar\nu_e+p\rightarrow e^+ + n$.
For $B=10^{16}$ G assumed above, 
$\langle\sigma^{(1,B)}_{\nu_en}\rangle$ is at most
4\% smaller than 
$\langle\sigma_{\nu_en}^{(1)}\rangle(1+\epsilon_{\chi_n})$ 
or
$\langle\sigma_{\nu_en}^{(1*)}\rangle(1+\epsilon_{\chi_n})$
while $\langle\sigma^{(1,B)}_{\bar\nu_ep}\rangle$
is indistinguishable from
$\langle\sigma_{\bar\nu_ep}^{(1*)}\rangle(1+\epsilon_{\chi_p})$.
This is because with 
$\langle E_{\nu_{e}}\rangle =11\,\text{MeV}$ and 
$\langle E_{\bar{\nu}_{e}}\rangle =16\,\text{MeV}$,
the important energy range for determining the average cross sections
has $E_\nu>\sqrt{eB}\sim 8$ MeV, for which the effects of Landau
levels are small. As $\langle E_{\bar{\nu}_{e}}\rangle$
is substantially larger than $\langle E_{\nu_{e}}\rangle$,
the magnetic field affects
$\langle\sigma^{(1,B)}_{\bar\nu_ep}\rangle$ even less than
$\langle\sigma^{(1,B)}_{\nu_en}\rangle$.

The results for absorption of supernova neutrinos on nucleons
can be summarized as follows. Generally speaking,
one can use 
$\langle\sigma^{(1)}_{\nu N}\rangle-\langle\sigma^{(0)}_{\nu N}\rangle$
to estimate the corrections without magnetic fields, and use
$\langle\sigma^{(0,B)}_{\nu N}\rangle-\langle\sigma^{(0)}_{\nu N}\rangle$
to estimate the corrections due to magnetic fields. 
For $B\lesssim 10^{16}$ G, 
$\langle\sigma^{(1)}_{\nu_en}\rangle+
\left[\langle\sigma^{(0,B)}_{\nu_en}\rangle-\langle\sigma^{(0)}_{\nu_en}\rangle\right]$
is a good estimate of $\langle\sigma^{(1,B)}_{\nu_en}\rangle$ with
an accuracy of $\sim 1$\%. For the same field strength, the effects of
magnetic fields are not important for 
$\langle\sigma_{\bar\nu_ep}\rangle$, and 
$\langle\sigma_{\bar\nu_ep}^{(1*)}\rangle$
is a good estimate of
$\langle\sigma^{(1,B)}_{\bar\nu_ep}\rangle$
with an accuracy of $\sim 1$\%. Note that we have ignored radiative 
corrections (see, e.g., \cite{Kurylov:2003}) and the effect of the 
Coulomb field of the proton on the electron wave function (see, e.g.,
\cite{Strumia:2003}). These factors give corrections at the level
of $\sim 2$\% \cite{Strumia:2003}. To account for them, we suggest
calculating  $\langle\sigma^{(1)}_{\nu_en}\rangle$ and
$\langle\sigma_{\bar\nu_ep}^{(1*)}\rangle$ as in 
Ref.~\cite{Strumia:2003} and use the results in the above estimates
for $\langle\sigma^{(1,B)}_{\nu_en}\rangle$
and $\langle\sigma^{(1,B)}_{\bar\nu_ep}\rangle$.

\subsection{Neutrino emission}

The differential reaction rates with respect to $\cos\Theta_{\nu}$
for $e^- + p \rightarrow \nu_e + n$ and 
$e^+ + n \rightarrow \bar\nu_e + p$ in a magnetic field have been
derived to $\mathcal{O}(1/m_N)$ as
$\ud\lambda_{e^{-}p}^{(1,B)}/\ud\cos\Theta_{\nu}$ and
$\ud\lambda_{e^{+}n}^{(1,B)}/\ud\cos\Theta_{\nu}$, respectively,
in Sec.~\ref{sec:diff}. To $\mathcal{O}(1)$, the zeroth order in
$1/m_N$, the differential reaction rates are \cite{Duan:2004nc}
\begin{equation}
\frac{\ud\lambda_{eN}^{(0,B)}}{\ud\cos\Theta_{\nu}}
= \frac{eB}{2\pi^2} \sum_{n_e} g_{n_e}
\int_0^{\infty}\frac{\ud\Gamma_{eN}^{(0,B)}/\ud\cos\Theta_\nu}
{\exp[(E_e/T)\mp\eta_e]+1}\ud k_{ez},
\label{eq:diff-lambda-eN-0b}
\end{equation}
where
\begin{align}
\frac{\ud\Gamma_{eN}^{(0,B)}}{\ud \cos\Theta_\nu}
& = \frac{\Gamma_{eN}^{(0)}}{2}
\left[ 1 + 2 \chi \frac{(f \mp g) g}{f^{2} + 3 g^{2}}
\cos \Theta_{\nu } \right] 
\nonumber \\
& \phantom{=} + \delta_{n_e,0} \frac{\Gamma_{eN}^{(0)}}{2}
\left[ \frac{f^{2}-g^{2}}{f^{2} + 3 g^{2}} \cos \Theta_{\nu} +
2 \chi \frac{(f \pm g) g}{f^{2} + 3 g^{2}}\right],
\label{eq:diff-gamma-eN-0b}\\
\Gamma_{eN}^{(0)}&=\frac{G_F^2\cos^2\theta_C}{2\pi}(f^2+3g^2)
(E_e \mp \Delta)^2.
\label{eq:gamma-eN-0b}
\end{align}
In the above equations and elsewhere in this subsection,
the upper sign is for $e^- + p \rightarrow \nu_e + n$
and the lower sign is for $e^+ + n \rightarrow \bar\nu_e + p$.
In Eq.~\eqref{eq:diff-gamma-eN-0b}, $\delta_{n_e,0}$ is the Kronecker
delta. For comparison, in the absence of any magnetic field, 
the differential reaction rates to $\mathcal{O}(1)$ are
\begin{equation}
\frac{\ud\lambda_{eN}^{(0)}}{\ud\cos\Theta_{\nu}} = 
\int\frac{\Gamma_{eN}^{(0)}}{\exp[(E_e/T)\mp\eta_e]+1}
\frac{\ud^3 k_e}{(2\pi)^3}.
\label{eq:lambda-eN-0}
\end{equation}
We have also calculated 
$\ud\lambda_{eN}^{(1*)}/\ud\cos\Theta_{\nu}$ 
to $\mathcal{O}(1/m_N)$ using the prescription in 
Ref.~\cite{Strumia:2003}.
Note that both $\ud\lambda_{eN}^{(0)}/\ud\cos\Theta_{\nu}$ 
and $\ud\lambda_{eN}^{(1*)}/\ud\cos\Theta_{\nu}$ are independent
of $\cos\Theta_{\nu}$.

\begin{figure*}
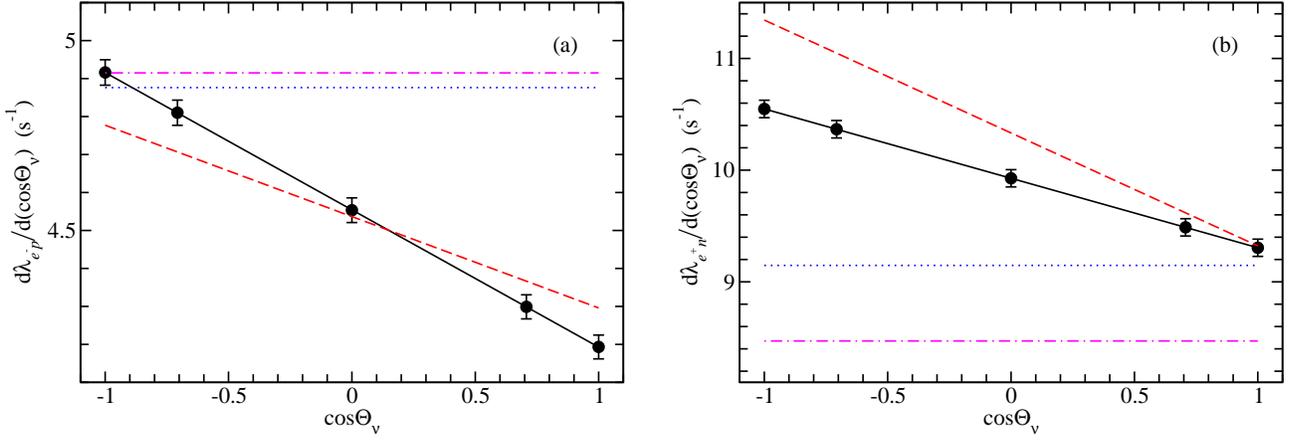

\begin{center}
$\begin{array}{@{}c@{\hspace{.2in}}c@{}}
\includegraphics*[width=3.25in, keepaspectratio]{fig2a.eps} &
\includegraphics*[width=3.25in, keepaspectratio]{fig2b.eps}
\end{array}$
\end{center}
\caption{\label{fig:eN-angle}%
(Color online) Comparison of various approximations for
$\ud\lambda_{eN}/\ud\cos\Theta_{\nu}$ [filled circles with
error bars: $\ud\lambda_{eN}^{(1,B)}/\ud\cos\Theta_{\nu}$;
dashed lines: $\ud\lambda_{eN}^{(0,B)}/\ud\cos\Theta_{\nu}$;
dotted lines: $\ud\lambda_{eN}^{(0)}/\ud\cos\Theta_{\nu}=
\lambda_{eN}^{(0)}/2$; dot-dashed lines:
$\ud\lambda_{eN}^{(1*)}/\ud\cos\Theta_{\nu}=
\lambda_{eN}^{(1*)}/2$]. The error bars on the filled
circles represent the accuracy of the numerical calculation
and the solid lines are linear fits to the filled circles.
These results are calculated using
$B=10^{16}\,\text{G}$, $T=2\,\text{MeV}$ for the matter temperature, 
and $\eta_e=0$.}
\end{figure*}

To compare $\ud\lambda_{eN}^{(1,B)}/\ud\cos\Theta_{\nu}$ with
$\ud\lambda_{eN}^{(0,B)}/\ud\cos\Theta_{\nu}$, we take
$B=10^{16}$ G, $T=2$ MeV, and $\eta_e=0$. The differential reaction
rates $\ud\lambda_{eN}^{(1,B)}/\ud\cos\Theta_{\nu}$ are numerically
calculated for
$\Theta_{\nu}=0$, $\pi/4$, $\pi/2$, $3\pi/4$, and $\pi$ and shown as
the filled circles with error bars in Fig.~\ref{fig:eN-angle}.
Here and elsewhere in this subsection, the error bars for our
results represent the accuracy of the numerical calculation.
To very good approximation, the rates 
$\ud\lambda_{eN}^{(1,B)}/\ud\cos\Theta_{\nu}$ are linear functions
of $\cos\Theta_{\nu}$ as shown by the solid lines fitted to the
numerical results in Fig.~\ref{fig:eN-angle}. The rates 
$\ud\lambda_{eN}^{(0,B)}/\ud\cos\Theta_{\nu}$ as
functions of $\cos\Theta_{\nu}$ are shown as the dashed lines
in the same figure. It can be seen that
relative to $\ud\lambda_{e^-p}^{(0,B)}/\ud\cos\Theta_{\nu}$,
$\ud\lambda_{e^-p}^{(1,B)}/\ud\cos\Theta_{\nu}$ is smaller
for $\cos\Theta_{\nu}\gtrsim 0.15$ but larger 
for $\cos\Theta_{\nu}<0.15$ due to corrections of 
$\mathcal{O}(1/m_N)$. So we expect that when integrated over
$\cos\Theta_{\nu}$, the difference between $\lambda_{e^-p}^{(0,B)}$
and $\lambda_{e^-p}^{(1,B)}$ is small. In contrast, corrections of 
$\mathcal{O}(1/m_N)$ make 
$\ud\lambda_{e^+n}^{(1,B)}/\ud\cos\Theta_{\nu}$ smaller
than $\ud\lambda_{e^+n}^{(0,B)}/\ud\cos\Theta_{\nu}$ for
all values of $\cos\Theta_{\nu}$. As discussed in the case of
neutrino absorption, such corrections are due to the effects of
weak magnetism and recoil of the final-state nucleons, which tend
to affect the processes involving $\bar\nu_e$ more than those
involving $\nu_e$. These corrections can also be seen by comparing
the zero-field results $\ud\lambda_{eN}^{(0)}/\ud\cos\Theta_{\nu}$
and $\ud\lambda_{eN}^{(1*)}/\ud\cos\Theta_{\nu}$, which are shown 
as the dotted and dot-dashed lines, respectively, in 
Fig.~\ref{fig:eN-angle}. Note that for the parameters adopted above,
the magnetic field decreases the rates for
$e^- + p \rightarrow \nu_e + n$ but increases those for
$e^+ + n \rightarrow \bar\nu_e + p$.

\begin{figure*}
\begin{center}
\includegraphics*[width=1.0\textwidth, keepaspectratio]{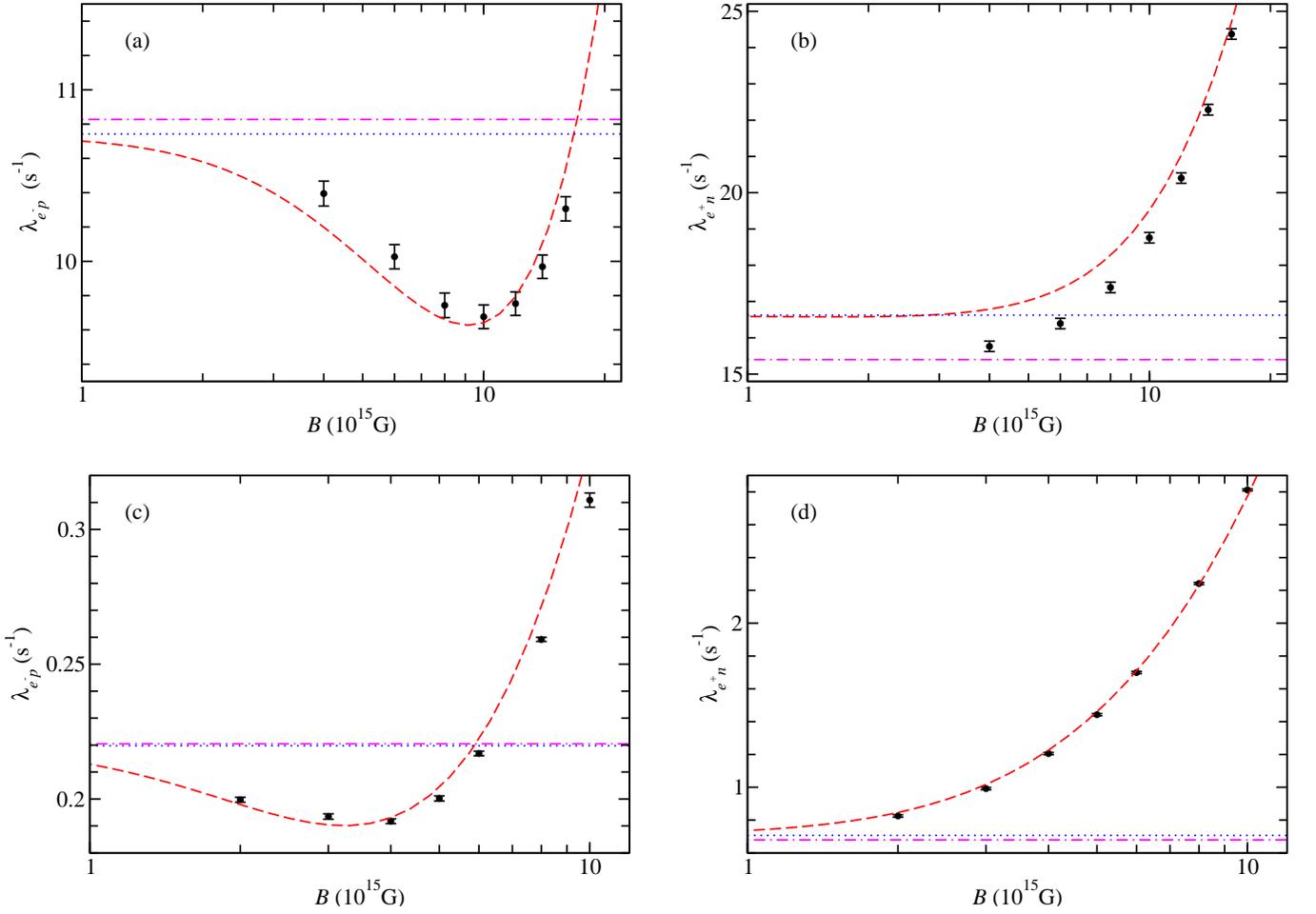}
\end{center}
\caption{\label{fig:eN-B}%
(Color online) Comparison of various approximations for
$\lambda_{eN}$ [filled circles with error bars:
$\lambda_{eN}^{(1,B)}$; dashed curves: $\lambda_{eN}^{(0,B)}$;
dotted lines: $\lambda_{eN}^{(0)}$; dot-dashed lines:
$\lambda_{eN}^{(1*)}$]. These results are calculated using
two sets of conditions: $(T,\ S,\ Y_e)=(2\ {\rm MeV},\ 50,\ 0.5)$
for (a) and (b) and $(1\ {\rm MeV},\ 100,\ 0.5)$ for (c) and (d),
where $T$ is the matter temperature.
The error bars on the filled circles represent the accuracy of 
the numerical calculation.}
\end{figure*}

To further explore the effects of the magnetic field on the
rates of neutrino emission, we consider two representative sets
of supernova conditions: $(T,\ S,\ Y_e)=(2\ {\rm MeV},\ 50,\ 0.5)$
and $(1\ {\rm MeV},\ 100,\ 0.5)$ for cases I and II, respectively.
For each case, the electron degeneracy parameter $\eta_e$ can be 
obtained from the equations of state as discussed in 
Ref.~\cite{Duan:2004nc}. We calculate the rates $\lambda_{eN}^{(1,B)}$
for a number of values of $B$ 
($4\times 10^{15}$--$1.6\times 10^{16}$ G for case I and
$2\times 10^{15}$--$10^{16}$ G for case II) and show the results as
the filled circles with error bars in Fig.~\ref{fig:eN-B}. The
corresponding rates $\lambda_{eN}^{(0,B)}$ as functions of $B$ and
the zero-field results $\lambda_{eN}^{(0)}$ and $\lambda_{eN}^{(1*)}$ 
are shown as the dashed curves and dotted and dot-dashed lines, 
respectively, in the same figure. As can be seen
from Fig.~\ref{fig:eN-B}, in the limit of small $B$, the dashed curves 
for $\lambda_{eN}^{(0,B)}$ agree with the dotted curves for 
$\lambda_{eN}^{(0)}$. The approach of $\lambda_{e^+n}^{(1,B)}$ 
(filled circles) to the zero-field limit $\lambda_{e^+n}^{(1*)}$
(dot-dashed line) is also clearly demonstrated for case I.
As a large number of Landau levels must be included in the
calculation for small $B$, it becomes computationally prohibitive
to demonstrate the behavior of $\lambda_{eN}^{(1,B)}$ in this limit
to the fullest extent. Nevertheless, the relation between
$\lambda_{eN}^{(1,B)}$ and $\lambda_{eN}^{(0,B)}$ for small $B$
clearly agrees with that between $\lambda_{eN}^{(1*)}$ and
$\lambda_{eN}^{(0)}$.

The dependences on $B$ for the rates of neutrino emission shown 
in Fig.~\ref{fig:eN-B} require discussion. The effects of the
magnetic field on these rates have been noted for the specific
case of $B=10^{16}$ G, $T=2$ MeV, and $\eta_e=0$ shown in
Fig.~\ref{fig:eN-angle}. More generally, Fig.~\ref{fig:eN-B} shows 
that $\lambda_{e^-p}^{(0,B)}$ and $\lambda_{e^-p}^{(1,B)}$ first
decrease with increasing $B$ to reach some minimum values and 
then increase with $B$. In contrast, $\lambda_{e^+n}^{(0,B)}$ and 
$\lambda_{e^+n}^{(1,B)}$ appear to increase monotonically with $B$.
The above results can be understood by 
considering two different effects of the magnetic field on 
$e^\pm$. On the one hand, a stronger field confines more $e^\pm$ 
to the ground Landau level, thus reducing the average $e^\pm$ 
energy. This tends to decrease the rates of neutrino emission. 
On the other hand, a magnetic field changes the $e^\pm$ phase 
space according to
\begin{equation}
2\int\frac{\ud^{3}k_{e}}{(2\pi)^{3}}
\longrightarrow
\frac{eB}{4\pi^{2}} \sum_{n_{e}} g_{n_{e}}
\int_{-\infty}^{+\infty} \ud k_{ez}.
\label{eq:phase-space}
\end{equation}
Thus, the $e^\pm$ phase space increases with $B$, which tends to
increase the rates of neutrino emission due to the increase in
the number density of $e^\pm$. The competition between the above 
two factors then determines the dependences on $B$ for the rates
of neutrino emission.

\begin{figure*}
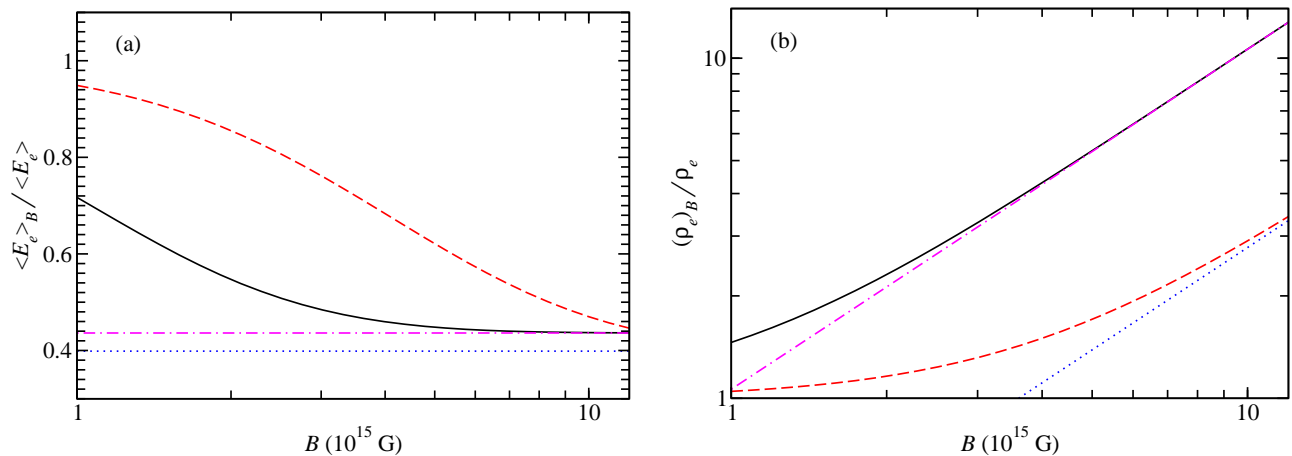

\begin{center}
$\begin{array}{@{}c@{\hspace{.2in}}c@{}}
\includegraphics*[width=3.25in, keepaspectratio]{fig4a.eps} &
\includegraphics*[width=3.25in, keepaspectratio]{fig4b.eps}
\end{array}$
\end{center}
\caption{\label{fig:B-effects}%
(Color online)
The ratios $\langle E_e\rangle_B/\langle E_e\rangle$ (a) and 
$(\rho_e)_B/\rho_e$ (b) as functions of $B$ (solid curves:
$T=1\,\text{MeV}$ for the matter temperature; dashed curves: 
$T=2\,\text{MeV}$). The
limiting case where all $e^\pm$ are in the ground Landau level
is shown as the dot-dashed ($T=1\,\text{MeV}$) and dotted
($T=2\,\text{MeV}$) lines. These results are calculated using
$\eta_e=0$ for simplicity.}
\end{figure*}

To show quantitatively the two effects of the magnetic field on
$e^\pm$ discussed above, we compare the average energy 
$\langle E_e\rangle_B$ and the number density $(\rho_e)_B$ of $e^\pm$
in a field with the corresponding quantities for no field,
$\langle E_e\rangle$ and $\rho_e$, respectively, for a wide range of
$B$ in Fig.~\ref{fig:B-effects}. As $0<\eta_e\ll 1$ for the supernova
conditions represented by cases I and II, we take $\eta_e=0$ for
simplicity. The major difference between these two cases lies in the
temperature. The ratios $\langle E_e\rangle_B/\langle E_e\rangle$
as functions of $B$ for $T=1$ and 2 MeV (cases II and I)
are shown as the solid and 
dashed curves, respectively, in Fig.~\ref{fig:B-effects}a. The
corresponding ratios $(\rho_e)_B/\rho_e$ are shown in 
Fig.~\ref{fig:B-effects}b. For large $B$, it is appropriate to consider
the limiting case where all $e^\pm$ are in the ground Landau level,
and therefore, $\langle E_e\rangle_B/\langle E_e\rangle$ is a
constant and $(\rho_e)_B/\rho_e$ increases linearly with $B$.
These limits are shown as the dot-dashed and dotted lines for
$T=1$ and 2 MeV, respectively, in Fig.~\ref{fig:B-effects}. As can be
seen from this figure, $\langle E_e\rangle_B/\langle E_e\rangle$
monotonically decreases with increasing $B$, eventually approaching 
the constant limit, while $(\rho_e)_B/\rho_e$ monotonically increases 
with $B$, eventually approaching the limiting linear trend.
The combined result of the two effects is that
$\lambda_{eN}$ decreases with increasing $B$ in weak field
regime, and starts to increase in strong field regime after
some turn-over point.  From dimensional analysis, we expect
the field at the turn-over point to be 
$B_\textrm{c}\sim E_\textrm{eff}^2/e$ with $E_\textrm{eff}$
 being some typical energy of the particles participating
in the reaction.
Because of the threshold, $e^-$ participating in
$e^- + p \rightarrow \nu_e + n$ is more energetic
than $e^+$ in $e^+ + n\rightarrow\bar\nu_e + p$.
So $B_\textrm{c}$ is larger for panels (a) and (c)
than for panels (b) and (d), respectively, in 
Fig.~\ref{fig:eN-B}. The turn-over points 
correspond to $B_\mathrm{c}\sim 2\times 10^{15}\ \mathrm{G}$
in panel (b) and $B_\mathrm{c}<10^{15}\ \mathrm{G}$ in panel (d).
However, the turn-over in these two panels is much weaker
than that in panels (a) and (c) so that $\lambda_{e^+ n}^{(0,B)}$
and $\lambda_{e^+ n}^{(1,B)}$ appear to increase monotonically
with $B$ for $B\gtrsim 10^{15}\ \mathrm{G}$.  In addition,
because $E_\textrm{eff}$ is higher for higher $T$, $B_\textrm{c}$
is larger for panels (a) and (b) ($T=2\ \mathrm{MeV}$)
than for panels (c) and (d) ($T=1\ \mathrm{MeV}$) in
 Fig.~\ref{fig:eN-B}.

In summary, we note that the rates $\lambda_{eN}$ are 
sensitive to the temperature $T$ of the supernova environment
regardless of $B$:
lowering $T$ by a factor of two reduces $\lambda_{e^-p}$ by factors
of $\sim 30$--50 and $\lambda_{e^+n}$ by factors of $\sim 6$--20
(see Fig.~\ref{fig:eN-B}). In contrast, the average cross sections
$\langle\sigma_{\nu N}\rangle$ only have minor dependence on $T$
(mainly through the polarization of nucleon spin) so that the
rates $\lambda_{\nu N}$ essentially scale
with the radius $R$ as $\lambda_{\nu N}\propto R^{-2}$. For the
temperature profile in the supernova environment of interest,
the rates $\lambda_{\nu N}$ dominate $\lambda_{eN}$. Therefore, 
so long as the former rates are calculated accurately, the latter
can be estimated using $\lambda_{eN}^{(0,B)}$ to good approximation 
for $B\lesssim 10^{16}$~G.

\section{Conclusions\label{sec:conclusions}}

In a previous paper \cite{Duan:2004nc}, we calculated the rates of 
$\nu_e + n \rightleftharpoons e^- + p$ and
$\bar\nu_e + p \rightleftharpoons e^+ + n$ in supernova environments
with strong magnetic fields assuming that the nucleon mass $m_N$ 
is infinite. We also applied these rates to discuss the implications 
of such fields for supernova dynamics. In the present paper, we
have taken into account the effects of a finite $m_N$ and
developed a numerical method for calculating the above rates to
$\mathcal{O}(1/m_N)$ for similar environments. Rates with such an 
accuracy are required for application to supernova nucleosynthesis.

We have shown that our 
results have the correct behavior in the limit of high neutrino
energy or small magnetic field. We find that for typical supernova
$\nu_e$ energy distributions, magnetic fields of $B\sim 10^{16}$~G
reduce the rate of $\nu_e + n \rightarrow e^- + p$ while the
$\mathcal{O}(1/m_N)$ corrections due to weak magnetism and nucleon 
recoil increase this rate. These two opposite effects tend to
cancel. On the other hand, the reduction of the rate of
$\bar\nu_e + p \rightarrow e^+ + n$ by the $\mathcal{O}(1/m_N)$ 
corrections dominates the magnetic field effects for
$B\lesssim 10^{16}$~G and typical supernova $\bar\nu_e$ energy 
distributions. We also find that for typical supernova conditions
relevant for heavy element nucleosynthesis, the rates of
$e^- + p \rightarrow \nu_e + n$ and $e^+ + n \rightarrow \bar\nu_e + p$
first decrease and then increase with increasing $B$. 
As it is extremely time consuming to numerically calculate
to $\mathcal{O}(1/m_N)$ the rates for the above processes in 
strong magnetic fields, we recommend that for $B\lesssim 10^{16}$~G,
the following approximations be implemented in models of supernova
nucleosynthesis. For $\nu_e + n \rightarrow e^- + p$, it is simple
to calculate the average cross section including the magnetic
field effects but no $\mathcal{O}(1/m_N)$ corrections
[$\langle\sigma_{\nu_en}^{(0,B)}\rangle$ in 
Table~\ref{tab:sigma-nuN-avg}] or vice versa
[$\langle\sigma_{\nu_en}^{(1)}\rangle$ in 
Table~\ref{tab:sigma-nuN-avg}]. By comparing the two with
$\langle\sigma_{\nu_en}^{(0)}\rangle$, one can estimate the effects of
magnetic fields and the $\mathcal{O}(1/m_N)$ corrections,
respectively. With these two kinds of corrections combined,
$\langle\sigma_{\nu_en}^{(1)}\rangle+
\left[\langle\sigma_{\nu_en}^{(0,B)}\rangle-\langle\sigma_{\nu_en}^{(0)}\rangle\right]$
agrees with the result of the full calculation
at the level of $\sim 1$\%. 
For $\bar\nu_e + p \rightarrow e^+ + n$,
the magnetic field effects on the average cross section can be
ignored but the $\mathcal{O}(1/m_N)$ corrections should be
included with an exact treatment of the reaction kinematics
[$\langle\sigma_{\bar\nu_ep}^{(1*)}\rangle$ 
in Table~\ref{tab:sigma-nuN-avg}]. While we have ignored 
radiative corrections and the effect of the Coulomb field of the proton
on the electron wave function, these factors can be included in
$\langle\sigma_{\nu_en}^{(1)}\rangle$ and
$\langle\sigma_{\bar\nu_ep}^{(1*)}\rangle$ following 
Ref.~\cite{Strumia:2003}. For 
$e^- + p \rightarrow \nu_e + n$ and 
$e^+ + n \rightarrow \bar\nu_e + p$, the rates including the
magnetic field effects but no $\mathcal{O}(1/m_N)$ corrections
[Eqs.~(\ref{eq:diff-lambda-eN-0b})--(\ref{eq:gamma-eN-0b})] are 
sufficient.

In conclusion, we note that the cross sections of neutrino absorption on
nucleons are relevant not only for supernova nucleosynthesis but also
for determining the thermal decoupling of $\nu_e$ and $\bar\nu_e$
from the protoneutron star. For example, a decrease in
$\sigma_{\bar\nu_ep}$ would enable $\bar\nu_e$ to emerge from
deeper and hotter regions of the protoneutron star, thus increasing
the average $\bar\nu_e$ energy. Accurate neutrino energy spectra are
essential to models of supernova nucleosynthesis \cite{Qian:1996}. 
However, our results on the cross sections for neutrino absorption 
cannot be applied directly to the discussion of neutrino decoupling 
from a strongly-magnetized protoneutron star because the conditions
(e.g., temperature and density) inside such
a star are very different from those considered here. We also note
that neutrino scattering on $e^\pm$ plays a significant role in 
supernova explosion \cite{Bethe:1990mw}.
Similar to the case of $e^\pm$ capture on nucleons, the rates 
of neutrino scattering on $e^\pm$ above the protoneutron star will
be modified substantially by strong magnetic fields. These issues
remain to be explored in detail by future studies.

\begin{acknowledgments}
We would like to thank Arkady Vainshtein for helpful discussions. 
We also want to thank John Beacom for communications regarding the
effects of weak magnetism and nucleon recoil on neutrino processes. 
This work was supported in part by DOE grant DE-FG02-87ER40328.
\end{acknowledgments}

\appendix

\section{Special Function%
\label{sec:algorithm-of-I}}

The special function $I_{n,r}(\zeta)$ can be written as \citep{Arras:1998mv}
\begin{equation}
I_{n,r}(\zeta)=\sqrt{\frac{r!}{n!}}
e^{-\zeta/2}\zeta^{(n-r)/2}L_{r}^{n-r}(\zeta),
\end{equation}
where $L_{n}^{\alpha}(x)$ is the generalized Laguerre polynomial
defined as \citep{Gradshteyn:1980}
\begin{subequations}
\begin{eqnarray}
L_{n}^{\alpha}(x) & = & 
\frac{1}{n!}e^{x}x^{-\alpha}\frac{\ud^{n}}{\ud x^{n}}(e^{-x}x^{n+\alpha})\\
 & = & \sum_{m=0}^{n}(-1)^{m}\left(\begin{array}{c}
n+\alpha\\
n-m\end{array}\right)\frac{x^{m}}{m!}.
\end{eqnarray}
\end{subequations}
To calculate $I_{n,r}(\zeta)$ efficiently, we use its properties 
given below.

\begin{itemize}
\item \emph{Mirror relation}\\
Based on the identity \citep{Sokolov:1968}
\begin{equation}
(-1)^{n-r}\zeta^{-(n-r)}Q_{n}^{r-n}(\zeta)=Q_{r}^{n-r}(\zeta),
\end{equation}
 where 
\begin{equation}
Q_{r}^{n-r}(\zeta)\equiv r!L_{r}^{n-r}(\zeta),
\end{equation}
 it is straightforward to show that 
\begin{equation}
I_{n,r}(\zeta)=(-1)^{n-r}I_{r,n}(\zeta).
\label{eq:I-mirror-rel}
\end{equation}

\item \emph{Recursion relation}s\\
 Using the recursion relation of the generalized Laguerre polynomial
\citep{Gradshteyn:1980}
\begin{equation}
L_{n}^{\alpha-1}(\zeta)=L_{n}^{\alpha}(\zeta)-L_{n-1}^{\alpha}(\zeta),
\end{equation}
 one can show that 
\begin{eqnarray}
I_{n,r}(\zeta) & = & \sqrt{\frac{r!}{n!}}
e^{-\zeta/2}\zeta^{(n-r)/2}L_{r}^{n-r}(\zeta)
\left[L_{r}^{(n-1)-r}(\zeta)+L_{r-1}^{(n-1)-(r-1)}(\zeta)\right]\nonumber \\
 & = & \sqrt{\frac{\zeta}{n}}I_{n-1,r}(\zeta)
+\sqrt{\frac{r}{n}}I_{n-1,r-1}(\zeta).
\label{eq:I-recursion-rel-1}
\end{eqnarray}
 Using this recursion relation and the mirror relation 
in Eq.~\eqref{eq:I-mirror-rel},
one can prove another recursion relation 
\begin{equation}
I_{n,r}(\zeta)=-\sqrt{\frac{\zeta}{r}}I_{n,r-1}(\zeta)
+\sqrt{\frac{n}{r}}I_{n-1,r-1}(\zeta).
\label{eq:I-recursion-rel-2}
\end{equation}
\end{itemize}

Starting from the definition 
\begin{equation}
I_{0,0}(\zeta)=e^{-\zeta/2},
\end{equation}
one can use the recursion relation in Eq.~\eqref{eq:I-recursion-rel-1}
to obtain
\begin{equation}
I_{n,0}(\zeta)=\sqrt{\frac{\zeta^{n}}{n!}}I_{0,0}(\zeta).
\label{eq:I-n0}
\end{equation}
Using the above result and the mirror relation in 
Eq.~\eqref{eq:I-mirror-rel}, one has 
\begin{equation}
I_{0,r}(\zeta)=(-1)^{r}\sqrt{\frac{\zeta^{n}}{r!}}I_{0,0}(\zeta).
\end{equation}
The function $I_{n,r}(\zeta)$ with $n>0$ and $r>0$ can be
calculated as follows:
\begin{enumerate}
\item Compute $I_{n-1,0}(\zeta)$ and $I_{n,0}(\zeta)$ from 
Eq.~\eqref{eq:I-n0}. Set $r^{\prime}=1$.
\item \label{step:mark}Compute $I_{n,r^{\prime}}(\zeta)$ from 
the recursion relation in Eq.~\eqref{eq:I-recursion-rel-2}.
\item If $r^{\prime}=r$, finish. Otherwise, compute 
$I_{n-1,r^{\prime}}(\zeta)$ from the recursion relation in
Eq.~\eqref{eq:I-recursion-rel-1}.
\item Advance $r^{\prime}$ by unity and
return to step~\ref{step:mark}.
\end{enumerate}

\section{Reduced Amplitude squared%
\label{sec:amp-sqr}}

The reduced amplitude squared $\mathcal{W}_{\nu_{e}n}$ for
$\nu_e + n \rightarrow e^- + p$ is defined in Eq.~\eqref{eq:W-nun-def}.
The amplitude $\mathfrak{M}_{\nu_{e}n}$ [Eq.~\eqref{eq:nun-amp}]
contained in $\mathcal{W}_{\nu_{e}n}$ can be simplified using
\begin{eqnarray}
 &  & \int_{0}^{\infty}\xi\,\ud\xi\int_{0}^{2\pi}
  e^{i\mathbf{w}_{\perp} \cdot \mathbf{x}_{\perp} %
-i(n_{e}-r_{e}-n_{p}+r_{p})\phi}
  I_{n_{p},r_{p}}(eB \xi^{2}/2)
  I_{n_{e},r_{e}}(eB \xi^{2}/2)
  \,\ud\phi\nonumber \\
 & = & \frac{2\pi}{eB}
  i^{(n_{e}-r_{e}-n_{p}+r_{p})}
  e^{-i(n_{e}-r_{e}-n_{p}+r_{p})\phi_{w}}
  I_{n_{e},n_{p}}(w_{\perp}^{2}/2eB)
  I_{r_{e},r_{p}}(w_{\perp}^{2}/2eB),
\label{eq:I-integral-rel}
\end{eqnarray}
where $\phi_{w}$ is the azimuthal angle of $\mathbf{w}_{\perp}$. 
The above result follows from 
\cite{Sokolov:1968,Gradshteyn:1980,Duan:2004nc}
\begin{equation}
\int_{0}^{2\pi} e^{i\mathbf{w}_{\perp}\cdot\mathbf{x}_{\perp}-i(n-r)\phi}
\,\ud\phi
 = 2\pi i^{n-r}e^{-i(n-r)\phi_{w}}
  J_{n-r}(w_{\perp}\xi)
\end{equation}
and
\begin{equation}
\int_{0}^{\infty} 
J_{(n-r)-(n^{\prime}-r^{\prime})}(2\sqrt{u\zeta})
I_{n^{\prime},r^{\prime}}(u)
I_{n,r}(u)\,\ud u
 = I_{n,n^{\prime}}(\zeta) I_{r,r^{\prime}}(\zeta),
\end{equation}
where $J_n(\zeta)$ is the Bessel function.

Noting that \cite{Sokolov:1968}
\begin{equation}
\sum_r I_{n,r}(\zeta) I_{n^\prime,r}(\zeta) = \delta_{n,n^\prime}
\end{equation}
and using Eqs.~\eqref{eq:degeneracy-r} and \eqref{eq:I-integral-rel},
we are able to derive the following explicit expressions of 
$\mathcal{W}_{\nu_{e}n}$ with the help of 
\textsc{Mathmatica}\textsuperscript{\textregistered}:
\begin{subequations}\label{eq:W-nun}
\allowdisplaybreaks
\begin{eqnarray}
(\mathcal{W}_{\nu_{e}n})_{s_{p}=1,s_{n}=1} & = & 
(f+g)^{2} (1+v_{ez}) (1+\cos\Theta_{\nu})
I_{n_{e},n_{p}}^{2}(w_{\perp}^{2}/2eB)\nonumber \\
 &  & + (f-g)^{2} (1-v_{ez}) (1-\cos\Theta_{\nu})
I_{n_{e}-1,n_{p}}^{2}(w_{\perp}^{2}/2eB)\nonumber \\
 &  & + 2(f^{2}-g^{2}) \frac{\sqrt{2n_{e}eB}}{E_{e}}
\cos\phi_{w} \sin\Theta_{\nu}
I_{n_{e}-1,n_{p}}(w_{\perp}^{2}/2eB)
I_{n_{e},n_{p}}(w_{\perp}^{2}/2eB)\nonumber \\
 &  & + \frac{1}{m_{N}} \Big\{
\Big[-(f+g)^{2} (1+v_{ez}) (1+\cos\Theta_{\nu}) (k_{nz}+k_{pz})\nonumber \\
 &  & \hphantom{+\frac{1}{m_{N}}\Big\{\Big[}
 - (f+g) (2f+f_{2}) (1+v_{ez}) \sin\Theta_{\nu} k_{nx}\nonumber \\
 &  & \hphantom{+\frac{1}{m_{N}}\Big\{\Big[}
 + f_{2}(f+g) (1+v_{ez}) \sin\Theta_{\nu} w_{x}\nonumber \\
 &  & \hphantom{+\frac{1}{m_{N}}\Big\{\Big[}
 + (f+g) (2f+f_{2}) (1+\cos\Theta_{\nu}) \frac{2n_{e}eB}{E_{e}}
\Big] \times I_{n_{e},n_{p}}^{2}(w_{\perp}^{2}/2eB)\nonumber \\
 &  & \hphantom{+\frac{1}{m_{N}}\Big\{} + 
\Big[(f-g)^{2} (1-v_{ez}) (1-\cos\Theta_{\nu}) (k_{nz}+k_{pz})\nonumber \\
 &  & \hphantom{+\frac{1}{m_{N}}\Big\{+\Big[}
 + f_{2}(f-g) (1-v_{ez}) \sin\Theta_{\nu} k_{nx}\nonumber \\
 &  & \hphantom{+\frac{1}{m_{N}}\Big\{+\Big[}
 - (f-g)(2f+f_{2}) (1-v_{ez}) \sin\Theta_{\nu} w_{x}\nonumber \\
 &  & \hphantom{+\frac{1}{m_{N}}\Big\{+\Big[}
 - f_{2}(f-g) (1-\cos\Theta_{\nu}) \frac{2n_{e}eB}{E_{e}}
\Big] \times I_{n_{e}-1,n_{p}}^{2}(w_{\perp}^{2}/2eB)\nonumber \\
 &  & \hphantom{+\frac{1}{m_{N}}\Big\{} + 
\Big[-f_{2}(f+g) (1+v_{ez}) \sin\Theta_{\nu} 
  \cos\phi_{w}\sqrt{2n_{e}eB}\nonumber \\
 &  & \hphantom{+\frac{1}{m_{N}}\Big\{+\Big[}
 + (f-g)(2f+f_{2}) (1-v_{ez}) \sin\Theta_{\nu} 
  \cos\phi_{w} \sqrt{2n_{e}eB}\nonumber \\
 &  & \hphantom{+\frac{1}{m_{N}}\Big\{+\Big[}
 + 2\bigl(-f^{2}+g(f+f_{2})+f(f-g+f_{2}) \cos\Theta_{\nu} \bigr)\nonumber \\
 &  & \hphantom{+\frac{1}{m_{N}}\Big\{+\Big[+}
  \times \cos(\Phi_{n}-\phi_{w}) 
  \frac{\sqrt{2n_{e}eB}k_{n\perp}}{E_{e}}\nonumber \\
 &  & \hphantom{+\frac{1}{m_{N}}\Big\{+\Big[}
 - (f+g)(2f+f_{2}) (1+\cos\Theta_{\nu}) 
  \frac{\sqrt{2n_{e}eB}w_{\perp}}{E_{e}}\nonumber \\
 &  & \hphantom{+\frac{1}{m_{N}}\Big\{+\Big[}
 + f_{2}(f-g) (1-\cos\Theta_{\nu})
  \frac{\sqrt{2n_{e}eB}w_{\perp}}{E_{e}}
\Big]\nonumber \\
 &  & \hphantom{+\frac{1}{m_{N}}\Big\{+}
\times I_{n_{e}-1,n_{p}}(w_{\perp}^{2}/2eB)
  I_{n_{e},n_{p}}(w_{\perp}^{2}/2eB)\Big\},
\end{eqnarray}
\begin{eqnarray}
(\mathcal{W}_{\nu_{e}n})_{s_{p}=1,s_{n}=-1} & = & 
4g^{2} (1+v_{ez}) (1-\cos\Theta_{\nu})
I_{n_{e},n_{p}}^{2}(w_{\perp}^{2}/2eB)\nonumber \\
 &  & + \frac{2}{m_{N}} \Big\{
\Big[-g(f+g+f_{2}) (1+v_{ez}) \sin\Theta_{\nu}k_{nx}\nonumber \\
 &  & \hphantom{+\frac{2}{m_{N}}\Big\{\Big[}
 + 2g(f+f_{2}) (1+v_{ez}) (1-\cos\Theta_{\nu}) (k_{ez}-k_{\nu z})\nonumber \\
 &  & \hphantom{+\frac{2}{m_{N}}\Big\{\Big[}
 + g(f+g+f_{2}) (1-\cos\Theta_{\nu}) \frac{2n_{e}eB}{E_{e}}
\Big] \times I_{n_{e},n_{p}}^{2}(w_{\perp}^{2}/2eB)\nonumber \\
 &  & \hphantom{+\frac{2}{m_{N}}\Big\{}
 + g(f-g+f_{2}) (1+v_{ez}) \sin\Theta_{\nu} w_{x}
  I_{n_{e},n_{p}-1}^{2}(w_{\perp}^{2}/2eB)\nonumber \\
 &  & \hphantom{+\frac{2}{m_{N}}\Big\{} +
\Big[g(f-g+f_{2}) k_{n\perp} \cos(\Phi_{n}-\phi_{w})
 - g(f+g+f_{2}) w_{\perp}
\Big]\nonumber \\
 &  & \hphantom{+\frac{2}{m_{N}}\Big\{+}
  \times(1-\cos\Theta_{\nu}) \frac{\sqrt{2n_{e}eB}}{E_{e}}
  I_{n_{e}-1,n_{p}}(w_{\perp}^{2}/2eB)
  I_{n_{e},n_{p}}(w_{\perp}^{2}/2eB)\nonumber \\
 &  & \hphantom{+\frac{2}{m_{N}}\Big\{}
 - g(f-g+f_{2}) (1+v_{ez}) \sin\Theta_{\nu} \cos\phi_{w}
  \sqrt{2n_{e}eB}\nonumber \\
 &  & \hphantom{+\frac{2}{m_{N}}\Big\{-}
  \times I_{n_{e}-1,n_{p}-1}(w_{\perp}^{2}/2eB)
  I_{n_{e},n_{p}-1}(w_{\perp}^{2}/2eB)
\Big\},
\end{eqnarray}
\begin{eqnarray}
(\mathcal{W}_{\nu_{e}n})_{s_{p}=-1,s_{n}=1} & = & 
4g^{2} (1-v_{ez}) (1+\cos\Theta_{\nu})
I_{n_{e}-1,n_{p}-1}^{2}(w_{\perp}^{2}/2eB)\nonumber \\
 &  & + \frac{2}{m_{N}} \Big\{
\Big[-g(f+g+f_{2}) (1-v_{ez}) \sin\Theta_{\nu}k_{nx}\nonumber \\
 &  & \hphantom{+\frac{2}{m_{N}}\Big\{\Big[}
 - 2g(f+f_{2}) (1-v_{ez}) (1+\cos\Theta_{\nu}) (k_{ez}-k_{\nu z})\nonumber \\
 &  & \hphantom{+\frac{2}{m_{N}}\Big\{\Big[}
 + g(f+g+f_{2}) (1+\cos\Theta_{\nu}) \frac{2n_{e}eB}{E_{e}}
\Big] \times I_{n_{e}-1,n_{p}-1}^{2}(w_{\perp}^{2}/2eB)\nonumber \\
 &  & \hphantom{+\frac{2}{m_{N}}\Big\{}
 + g(f-g+f_{2}) (1-v_{ez}) \sin\Theta_{\nu} w_{x}
  I_{n_{e}-1,n_{p}}^{2}(w_{\perp}^{2}/2eB)\nonumber \\
 &  & \hphantom{+\frac{2}{m_{N}}\Big\{} +
\Big[g(f-g+f_{2}) k_{n\perp} \cos(\Phi_{n}-\phi_{w})
 - g(f+g+f_{2}) w_{\perp}
\Big]\nonumber \\
 &  & \hphantom{+\frac{2}{m_{N}}\Big\{+}
  \times(1+\cos\Theta_{\nu}) \frac{\sqrt{2n_{e}eB}}{E_{e}}
  I_{n_{e}-1,n_{p}-1}(w_{\perp}^{2}/2eB)
  I_{n_{e},n_{p}-1}(w_{\perp}^{2}/2eB)\nonumber \\
 &  & \hphantom{+\frac{2}{m_{N}}\Big\{}
 - g(f-g+f_{2}) (1-v_{ez}) \sin\Theta_{\nu} \cos\phi_{w}
  \sqrt{2n_{e}eB}\nonumber \\
 &  & \hphantom{+\frac{2}{m_{N}}\Big\{-}
  \times I_{n_{e}-1,n_{p}}(w_{\perp}^{2}/2eB)
  I_{n_{e},n_{p}}(w_{\perp}^{2}/2eB)
\Big\},
\end{eqnarray}
\begin{eqnarray}
(\mathcal{W}_{\nu_{e}n})_{s_{p}=-1,s_{n}=-1} & = & 
  (f+g)^{2} (1-v_{ez}) (1-\cos\Theta_{\nu})
  I_{n_{e}-1,n_{p}-1}^{2}(w_{\perp}^{2}/2eB)\nonumber \\
 &  & 
  +(f-g)^{2} (1+v_{ez}) (1+\cos\Theta_{\nu})
  I_{n_{e},n_{p}-1}^{2}(w_{\perp}^{2}/2eB)\nonumber \\
 &  & + 2(f^{2}-g^{2})   \frac{\sqrt{2n_{e}eB}}{E_{e}}
  \cos\phi_{w} \sin\Theta_{\nu} 
  I_{n_{e}-1,n_{p}-1}(w_{\perp}^{2}/2eB)
  I_{n_{e},n_{p}-1}(w_{\perp}^{2}/2eB)\nonumber \\
 &  & + \frac{1}{m_{N}} \Big\{
\Big[(f+g)^{2} (1-v_{ez}) (1-\cos\Theta_{\nu}) (k_{nz}+k_{pz})\nonumber \\
 &  & \hphantom{+\frac{1}{m_{N}}\Big\{\Big[}
 - (f+g)(2f+f_{2}) (1-v_{ez}) \sin\Theta_{\nu} k_{nx}\nonumber \\
 &  & \hphantom{+\frac{1}{m_{N}}\Big\{\Big[}
 + f_{2}(f+g) (1-v_{ez}) \sin\Theta_{\nu} w_{x}\nonumber \\
 &  & \hphantom{+\frac{1}{m_{N}}\Big\{\Big[}
 + (f+g)(2f+f_{2}) (1-\cos\Theta_{\nu}) \frac{2n_{e}eB}{E_{e}}
\Big] \times I_{n_{e}-1,n_{p}-1}^{2}(w_{\perp}^{2}/2eB)\nonumber \\
 &  & \hphantom{+\frac{1}{m_{N}}\Big\{} + 
\Big[-(f-g)^{2} (1+v_{ez}) (1+\cos\Theta_{\nu}) (k_{nz}+k_{pz})\nonumber \\
 &  & \hphantom{+\frac{1}{m_{N}}\Big\{+\Big[}
 + f_{2}(f-g) (1+v_{ez}) \sin\Theta_{\nu} k_{nx}\nonumber \\
 &  & \hphantom{+\frac{1}{m_{N}}\Big\{+\Big[}
 - (f-g)(2f+f_{2}) (1+v_{ez}) \sin\Theta_{\nu} w_{x}\nonumber \\
 &  & \hphantom{+\frac{1}{m_{N}}\Big\{+\Big[}
 - f_{2}(f-g) (1+\cos\Theta_{\nu}) \frac{2n_{e}eB}{E_{e}}\Big]
  \times I_{n_{e},n_{p}-1}^{2}(w_{\perp}^{2}/2eB)\nonumber \\
 &  & \hphantom{+\frac{1}{m_{N}}\Big\{} + 
\Big[
   - f_{2}(f+g) (1-v_{ez}) \sin\Theta_{\nu} \cos\phi_{w} 
  \sqrt{2n_{e}eB}\nonumber \\
 &  & \hphantom{+\frac{1}{m_{N}}\Big\{+\Big[}
  +(f-g)(2f+f_{2}) (1+v_{ez}) \sin\Theta_{\nu} \cos\phi_{w}
  \sqrt{2n_{e}eB}\nonumber \\
 &  & \hphantom{+\frac{1}{m_{N}}\Big\{+\Big[}
 - 2\bigl(f^{2}-g(f+f_{2})+f(f-g+f_{2}) \cos\Theta_{\nu} \bigr)\nonumber \\
 &  & \hphantom{+\frac{1}{m_{N}}\Big\{+\Big[-}
  \times\cos(\Phi_{n}-\phi_{w}) 
  \frac{\sqrt{2n_{e}eB}k_{n\perp}}{E_{e}}\nonumber \\
 &  & \hphantom{+\frac{1}{m_{N}}\Big\{+\Big[}
 - (f+g)(2f+f_{2}) (1-\cos\Theta_{\nu})
  \frac{\sqrt{2n_{e}eB}w_{\perp}}{E_{e}}\nonumber \\
 &  & \hphantom{+\frac{1}{m_{N}}\Big\{+\Big[}
 + f_{2}(f-g) (1+\cos\Theta_{\nu}) \frac{\sqrt{2n_{e}eB}w_{\perp}}{E_{e}}
\Big]\nonumber \\
 &  & \hphantom{+\frac{1}{m_{N}}\Big\{+}
  \times I_{n_{e}-1,n_{p}-1}(w_{\perp}^{2}/2eB)
  I_{n_{e},n_{p}-1}(w_{\perp}^{2}/2eB)\Big\}.
\end{eqnarray}
\end{subequations} 
In the above equations, $v_{ez}=k_{ez}/E_e$. 
The reduced amplitude squared 
$\mathcal{W}_{\bar\nu_{e} p}$ for 
$\bar\nu_e + p \rightarrow e^+ + n$ can be obtained from these
equations by making the substitution given in
Eq.~\eqref{eq:xing-rel}.

\bibliography{ref}

\end{document}